\pdfoutput=1 

\RequirePackage{ifpdf}
\documentclass{JINST}
\usepackage{multirow}
\usepackage{ccaption}

\title{Use of sensor characterization data to tune electrostatic model parameters for LSST sensors}

\author{Andrew Rasmussen\\
SLAC National Accelerator Laboratory,\\
  2575 Sand Hill Rd., Menlo Park, CA 94024\\
E-mail: \email{arasmus@slac.stanford.edu}}

\abstract{We build on previous efforts to model CCD sensors, during illumination and collection of conversions. We use a finite summation of simple, electrostatic field models. The upgraded functionality of our framework provides specific predictions for perturbations in pixel boundary enclosures (e.g., at the backside window) and the bookkeeping capability to stack those perturbations so that they may be utilized as Greens functions -- portable calculation results that may be generically applied to a range of precision astronomy related problems that naturally including astrometric, photometric and shape transfer issues. We approach the topic of using ancillary pixel data, derived from the Greens function and the registered image, to analyze sky data and constrain object parameters of astronomical targets.}

\keywords{CCDs; pixel mapping; transverse drift; brighter-fatter effect; pixel covariance; instrument signature removal}

\begin{document}
\section{Introduction}\label{sec:a}

A fundamental assumption at work when we apply data from a photo-electronic recording device ({\it e.g.}, CCD) to precision astronomy needs is the notion that the geometric domain tied to each pixel can be quantified to some degree of precision. 
In two previous papers we describe drift calculations using finite sums of physically motivated electrostatic terms that obey boundary conditions\cite{Rasmussen_paccd_2014} and subsequent quantitative constraints of the electrostatic parameters using available characterization data\cite{Rasmussen_spie_2014}.

This paper provides an update on the electrostatic drift calculations using infinite grids and finite arrays of continuous- and truncated multipoles in arrangements that satisfy electrostatic boundary conditions. In \S\ref{sec:b} we briefly summarize work already described elsewhere, then in \S\ref{sec:c} we describe improvements in modeling and in the modularity of drift calculation results. We ultimately (\S\ref{sec:d}) provide some examples for how this framework can be utilized to routinely produce better interpretation of laboratory characterization data and how it can in turn be used to significantly reduce the systematics floor prior to discovery of any remaining effects, which are typically found only after large quantities of high quality exposures are available, acquired only under the best conditions.

\begin{figure}[tbp]
\centering
\includegraphics[width=.8\textwidth]{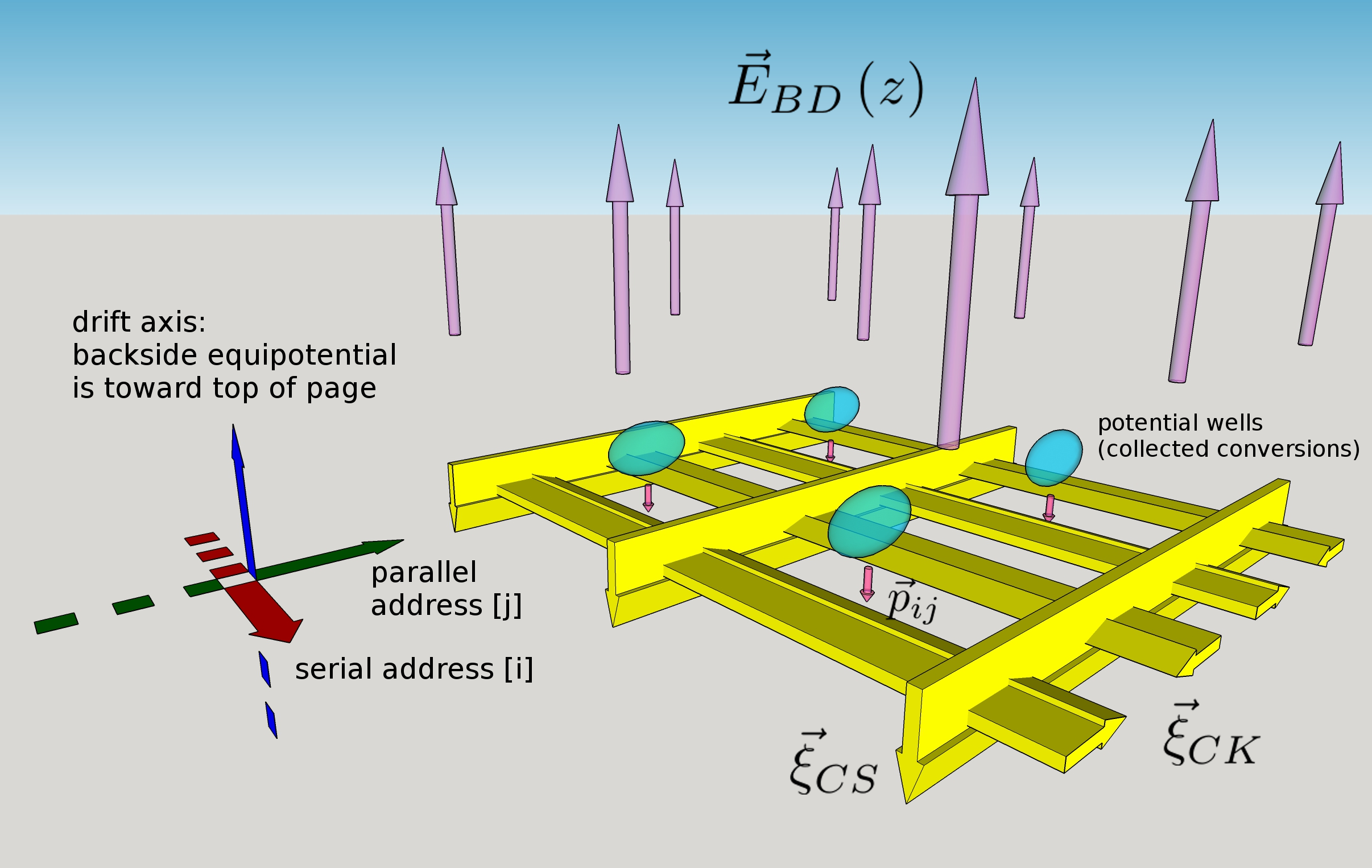}
\caption{Descriptions of elements contributing to the overall drift field $\vec{E}^{tot}(z)$ refer to items labeled in this figure. Distributed dipole moments with translational invariance along symmetry axes $\hat{s}$ are represented by the extruded arrows $\vec\xi_{CS}$ and $\vec\xi_{CK}$ which repeat every pixel; dipole moments induced by conversions collected at the channel are represented by the small arrows pointing downward $\vec{p}_{ij}$. The backdrop one-dimensional field $\vec{E}_{BD}$ is represented by the arrows pointing upward toward the backside electrode, which would be 10 pixel units away toward the top of the page. See text for more details.
}
\label{fig:sensor_geometry}
\end{figure}
 \section{Summary of prior work}\label{sec:b}

Figure~\ref{fig:sensor_geometry} is a perspective representation of a 2$\times$2 pixel section near the gate structure showing the electrostatic model elements used in the drift calculations. The distributed dipoles that act as barriers against conversion loss to adjacent pixels, are represented as extruded yellow arrows: The channel stop dipoles $\vec{\xi}_{CS}$ extend along the parallel address axis [j] while the clock dipoles $\vec{\xi}_{CK}$ lie in the plane of the front side gate structure and extend along the serial address axis [i]. When depleted, the p+ channel stop implants form localized, linear bound charge density distributions some distance away from the clocks: equipotential, polysilicon strips that pass underneath them in this picture. Since there can only be a normal electric field component $\vec{E}=(\vec{E}\cdot\hat{n})\cdot\hat{n}$ at the surfaces of those equipotentials, we model the electrostatic influence of the channel stops using an image charge distribution to form dipole moments $\vec{\xi}_{CS}$. Similar arguments are used to treat the influence of the periodic, alternating clock polarities as the in-plane dipole moments $\vec{\xi}_{CK}$. The zeroth order ``backdrop'' drift field $\vec{E}_{BD}(z)$ shown is responsible for drawing conversions from approximately 100$\mu$m (10 pixels) above the plane shown, down to the level of the buried channel (\cite[\S3]{Rasmussen_paccd_2014}). The combined effects from $\vec{E}_{BD}$ and the fields due to distributed dipole moments $\vec{\xi}_{CS}$ and $\vec{\xi}_{CK}$ -- give rise to the potential wells in 3 dimensions as well as to saddle point loci encountered by conversions that drift along $\vec{E}_{BD}$ close to the pixel boundaries ({\it e.g.}, \cite[Fig.1]{Rasmussen_spie_2014}). Finally, conversions collected and confined to the potential wells shown (as translucent ellipsoids) -- in turn set up image charge distributions across the polysilicon and an electrostatic influence characterized by pixel specific, dipole moments $\vec{p}_{ij}$. The following list enumerates details of, and deviations from, the nominal model described above -- in order to match and quantitatively compare with sensor response details routinely extracted from sensor characterization data. In all cases, pixel boundary distortions are computed by successively bisecting adjacent points that connect, via the sum total drift fields $\vec{E}^{tot}(\vec{x})$, to neighboring pixels according to methods previously described \cite[\S 4]{Rasmussen_paccd_2014}, \cite[\S 2 \it ff.]{Rasmussen_spie_2014}.

\begin{enumerate}

\item {\bf The backdrop field.} $\vec{E}_{BD}(z)$ and its relationship to clock rail voltages and backside bias $V_{BSS}$, was studied and constrained using available inferential methods. This was described in detail earlier~\cite[\S3.1]{Rasmussen_spie_2014} and comprises a tall pole in the physical description of the device under operation. 

\item {\bf Edge rolloff.} For the edges parallel to the channel stops, the regularly spaced arrangement of channel stops is truncated so that the outermost columns are bounded by channel stops, but with no more channel stop implants beyond. This abrupt change in the distribution of channel stops is sufficient to induce the drop in flat field response with a $\sim$10 pixel scale length and $\sim$1 pixel astrometric error \cite{Doherty_spie_2014,Rasmussen_spie_2014}. Presence of a guard ring electrode is modeled with a single distributed dipole moment $\vec{\xi}_{GD}$ in the plane of the gate structure, and positioned directly outside of the last channel stop implant. A negative voltage on the guard ring drain bias is modeled as $\vec{\xi}_{GD}$ pointing inward toward the imaging section of the sensor, and results in a modest improvement in photo response uniformity, with strongest response improvement in a small number of the outermost columns. See Figure~\ref{fig:edge-n-midline}, top row plots. 

\item {\bf Midline redistribution.} The bloom stop implant running perpendicular to the channel stops is modeled as a single, isolated implant with parameters identical to each of the channel stop implants. Its electrostatic influence is assumed to be identical to $\vec{\xi}_{CS}$ but with its symmetry axis parallel to the symmetry axes for $\vec{\xi}_{CK}$. Variation in the predicted, low light level flat field response is induced by adjusting the backside bias $V_{BSS}$. See Figure~\ref{fig:edge-n-midline}, bottom row plots.

\item \label{item:pixie} {\bf Dark columns at amplifier boundaries and bamboo.} Localized variations in barrier fields characterized by $\vec{\xi}_{CS_i}$ and/or $\vec{\xi}_{CK_j}$: 
Isolated~\cite[Fig.3, tearing onset]{Rasmussen_paccd_2014} and collective~\cite[Fig.6, bamboo]{Rasmussen_spie_2014} variations in barrier strengths have been shown to cause pixel geometry distortions as influence functions. We suggest that to first order, small lithographic errors that cause variations in the separation between parallel clocks result in very little change to the distributed dipole moments $\vec{\xi}_{CK_j}$, but that errors in the midpoint position, or the location of the dipole, result in a first order translation of the pixel boundary between adjacent rows.  We therefore expect anticorrelation in area errors for pixels within the same column but from neighboring rows -- this was observed by Smith \& Rammer~\cite{Smith:2008} for smoothing rates with slope $-1$ when computing the standard deviation for averaged "lines" (rows).  In contrast, distributed dipole moments $\vec{\xi}_{CS_i}$ should be susceptible to implant errors (ion dosage \& depth) on top of any lithographic errors. As we showed previously~\cite[Fig.3]{Rasmussen_paccd_2014}, variations in dipole strength induces flat field distortion patterns that are different from the pixel boundary jitter, and should not necessarily follow the same slope seen in the smoothing rate.

\item {\bf Brighter--fatter effect.} As indicated above and elsewhere~\cite{Antilogus_paccd_2014,Guyonnet_2015}, these pixel boundary distortion patterns are induced by the charge pattern already collected midway through the integration, and influence drift paths of subsequent conversions. The dipole moments $\vec{p}_{ij}$ are expected to have magnitudes equal to twice the product of the collected charge in the pixel and the depth of the buried channel, measured from the plane of the gate structure. The depth of the channel is not known {\it a priori}, but can be inferred from details of the photon transfer (mean--variance) curve using a set of predictive calculations. Our previous, strawman attempt to do this using a simplistic, {\it nearest--neighbor only} boundary distortion model \cite[\S 4.5, Fig.7]{Rasmussen_paccd_2014} would be superseded with our current model, because detailed pixel area distortions we model now carry enough information to adhere to basic conservation rules. See Figures~\ref{fig:bf_recordedpars} (upper left panel) and \ref{fig:bf_recordedpars2} (top row panels) for examples of how pixel boundary shifts are calculated in response to various levels of charge integration levels and its distribution.

\item {\bf PRNU.} In the course of computing pixel geometry response (at the backside surface) due to small geometric and electrostatic deviations from the regular, repeated grid depicted in Figure~\ref{fig:sensor_geometry}, we realized above (item \#\ref{item:pixie}) that the {\it influence functions} on pixel geometry due to lithographic or process geometry differences are significantly different in scale and symmetry, and these could potentially explain the marked difference in photon response non-uniformity (PRNU) smoothing rates~\cite{Smith:2008} and power spectra density slopes~\cite{Kotov:2010} along the parallel and serial address axes. We haven't any more details on this subject now, but plan to investigate further in the near future, because differences in expected area structure functions along the two address axes could be useful in removing degeneracy between pixel area and pixel quantum efficiency in any detailed decomposition of sensor flat field response.

\item {\bf Charge transfer efficiency.} None of the work described here pertains to CTE, but we anticipate needs in the future to properly simulate the effects of traps unexposed by the fat zero provided by sky background, that happen to have de-trapping times comparable to parallel transfer timescales. There is already evidence~\cite{Baumer:privcomm} that would suggest localized fluctuations in parallel transfer efficiency at medium light levels, and is perhaps not reflected in the extended pixel edge response ({\it EPER}) method of quantifying the sensor's CTE for device acceptance purposes. Such CTE mechanisms may have non-negligible impacts on PSF estimation for high signal-to-noise images and have yet to be quantified.
\end{enumerate}

\section{Improvements to the current modeling framework and recalibration}\label{sec:c}

In \S\ref{sec:b} we briefly described the geometry where we apply the individual electrostatic contributions to the total drift field $\vec{E}^{tot}$ ({\it cf.}~\cite[\S\S3-4]{Rasmussen_paccd_2014}). Since publication of the second paper~\cite{Rasmussen_spie_2014}, several important changes have been implemented, which are listed here:

\begin{enumerate}
\item {\bf Dipole moment ratio calibration against observed flat field covariance ratios.} Relative values of distributed dipole moments $\vec\xi_{CS}$ and $\vec\xi_{CK}$ that nominally reproduce observed ratios in covariance terms $\Delta \ln A_{01}/\Delta \ln A_{10} \sim 3$~\cite[Fig.7]{Rasmussen_spie_2014}, must be recomputed because of the new, in-plane arrangement of dipole moments $\vec\xi_{CK}$ with the alternating polarity shown in Figure~\ref{fig:sensor_geometry}.  Because of the alternating polarity shown, it could be argued that the barrier influence of the nearest two clock pairs may resemble that of a quadrupole moment instead of a dipole moment - so a different radial dependence would be in effect in the far field approximation. The updated parameter determination is shown in Figure~\ref{fig:template_constraints_geopars} and discussed at more depth in \S\ref{sec:e}.

\item {\bf Update to the method of images strategy.} Another side effect of the new dipole moment orientations for $\vec\xi_{CK}$, is that the plane of the gate structure is no longer be treated as an equipotential plane imparting its boundary condition on the electrostatic solution~\cite[Eqs.4.10 \& 4.12]{Rasmussen_paccd_2014}. The modification of those equations that is consistent with a {\it single} equipotential plane at the backside entrance window (for field contributors that depend on $\vec\xi_{CK}$) is to include only a single image for each charge configuration contributing to $\vec{E}^{tot}$ in place of the expansion given in those equations. Equation~4.12 would be altered to truncate the sum from $0\le \ell \le 1$. An updated representation of Equation~4.10 now includes a mapping of the image dipole moment:
\begin{eqnarray}
\delta\vec{E}_\perp(\vec{x}|\vec{x}_0)&=&{1 \over 2\pi\epsilon_0\epsilon_{Si}}\sum^{1}_{\ell =0}{2[\vec{\xi}_\ell\cdot\hat{r}_{\ell\perp}]\hat{r}_{\ell \perp}-\vec{\xi}_\ell\over r^2_{\ell\perp}},\\
\vec\xi_1 &=& 2 \left( \vec\xi_0 \cdot \hat{k} \right) \hat{k} - \vec\xi_0,\\
\vec{r}_{\ell \perp}&=&\vec{x}_\perp-(\vec{x}_{0\perp}+2\,\ell \,t_{Si}\,\hat{k}); \;\;\hat{r}_{\ell \perp}\equiv{\vec{r}_{\ell \perp}\over |\vec{r}_{\ell \perp}|}.
\end{eqnarray}

\item {\bf Influence of an infinite, periodic arrangement of barrier terms.} The regularly spaced, electrostatic model contributors, summed over the infinite grid -- is now evaluated using the following periodic expression we derived analytically:
\begin{eqnarray}
\sum^{\infty}_{m=-\infty} {2\,[\vec\xi\cdot\hat{r}_{m\perp}] \hat{r}_{m\perp}-\vec\xi \over r_{m\perp}^2}  &=& 
{2\pi^2 \over {\mathrm p}^2} 
{(1-\cos a \cosh b)\over \left(\cos a -\cosh b \right)^2 }\left([\vec\xi\cdot \hat{e}_\perp]\hat{e}_\perp - [\vec\xi \cdot\hat{k}]\hat{k}\right)\nonumber \\
&&+{2\pi^2 \over {\mathrm p}^2}{(\sin a \sinh b) \over (\cos a - \cosh b)^2}\left( [\vec\xi\cdot \hat{e}_\perp] \hat{k} + [\vec\xi\cdot \hat{k}] \hat{e}_\perp \right),\\
a&\equiv&2\pi |e_\perp |/\mathrm p,\\
b&\equiv&2\pi [\vec r \cdot \hat k] /\mathrm p,\\
\vec r &\equiv& \vec x - \vec x_0,\\
\vec r_\perp &\equiv& \vec r - [\vec r\cdot \hat{s}]\hat{s} = [\vec r \cdot \hat{k}]\hat{k} + [\vec r \cdot \hat{e}_\perp ]\hat{e}_\perp ,\\
\vec r_{m\perp}&\equiv&\vec{r}_\perp + {\mathrm m}\,{\mathrm p}\,\hat{e}_\perp,
\end{eqnarray} 
where $\hat s$ is the symmetry axis of the distributed dipole and $\hat k$ is the axis perpendicular to the plane of the grid of distributed dipoles, which are in turn spaced at regular intervals with (pixel) spacing $\mathrm p$ along the direction $\hat e_\perp$ in the gate structure plane. It follows that $(\hat k, \hat s, \hat e_\perp)$ form an orthonormal set of unit vectors. Use of this closed expression, together with a single image term per electrostatic elements $\vec\xi_{CK}$ rather than a finite expansion, greatly accelerates evaluation of the drift field within the pixel volume. Any deviation from the perfectly uniform and regular arrangement in distributed dipoles is easily be implemented by cancelation and modification of individual dipole terms in the expression, since the principle of superposition applies.

\item {\bf Discontinuities in the distributed dipoles.} Expressions were also derived to represent the influence of distributed dipole moments that are {\it truncated} -- that feature finite edge effect, field components parallel to the symmetry axis $\hat s$ nearby the truncation coordinate. These components are useful in modeling the effects of tearing - where apparently the hole concentration within the channel stops can abruptly transition from zero ({\it i.e.}, depleted) to some saturation level where transfer efficiencies drop nearly to zero. The superposition of such truncated, distributed dipole moments in adjacent columns can reproduce the sort of bimodal features seen in flat field response near the position of the change in hole concentration: this is the contour seen in flat field data where ``tearing'' is a problem ({\it cf.}~\cite[Fig.5]{Rasmussen_paccd_2014}). It also applies to the ``bamboo'' phenomenon ({\it cf.}~\cite[\S4.2]{Rasmussen_spie_2014}) for reasons described there. 

The expressions derived for the truncated, distributed dipole moment are:
\begin{eqnarray}
4\pi\epsilon_0\epsilon_{Si}r_\perp^2 \delta\vec{E}(\vec{x}|\vec{x}_0,\hat{s})
&=&\left(2-3\cos\theta_0+\cos^3\theta_0\right) [\vec\xi\cdot\hat{r}_\perp]\,\hat{r}_\perp\nonumber\\
&&+\sin^3\theta_0\,[\vec\xi\cdot\hat{r}_\perp]\hat{s} - (1-\cos \theta_0)\vec\xi,\\
\cos\theta_0&\equiv&\hat{s}\cdot\hat{r}.
\end{eqnarray}
Here, the distributed axis of symmetry unit vector $\hat{s}$ points in the direction of the truncation, {\it i.e.} from the truncation point $\vec{x}_0$ away from where the distributed charges are situated. As required, this expression converges to the familiar distributed dipole expression in the limiting cases. However, the truncated distributed dipole model does not lend well to forming concise expressions such as the infinite sum above, so a summation over a finite number of electrostatic components is performed when necessary for certain fixed pattern features ({\it e.g.}, discontinuous features associated with {\it tearing onset}~\cite[Fig.3, bottom row plots]{Rasmussen_paccd_2014}; ``bamboo'' features~\cite[Fig.6]{Rasmussen_spie_2014}).

\item {\bf Porting drift calculation results to modeling platforms.} We have recently devised a way to make drift calculation results portable so that contributions from various effects can be efficiently scaled and combined in downstream simulation programs -- where pixel boundaries are treated as simple polygons with $N$ ordered vertices ($x_k$,$y_k$) with $k=0..N-1$, closing them with $(x_N,y_N)\equiv(x_0,y_0)$. In the limit that drift calculations produce null results, the pixel boundaries then follow the usual {\it square contours}, consistent with the fundamental assumption of common pixel data pipeline routines. The information carried from the drift calculations include, for each segment of the pixel boundary, specific to a given aggressor offset vector and amplitude:
\begin{itemize}
\item Lateral, two dimensional displacements of the positions at the backside window that map to the corners of the registered pixel for that segment (4 numbers);
\item An array of normal displacements perpendicular to the segment. The number of samples, together with an array of the displacement values with uniform spacing along the nominal segment axis. 
\end{itemize}
Our downstream simulator loads these distortion patterns as {\it templates}, scales and superposes the effects in according to rules of the model, and efficiently evaluates the pixel boundaries. Boundaries represented as ordered polygon vertices are made accessible from the data structures or classes that represent each pixel. The polygonal pixel structure (together with higher level summary quantities) lends to efficient {\it point-in-polygon} algorithms to test domain membership, and to robust, single pass, geometric formulae. The following pixel moment expressions were derived by generalizing the shoelace formula for polygon area:
\begin{eqnarray}
\mathrm{A}_{ij}&=& +{1 \over 2} \sum_{k=0}^{N-1} (x_{k+1}y_{k}-x_{k}y_{k+1})\label{eq:aij}\\
\mathrm{I\textsc{xx}}_{ij}\mathrm A_{ij}&=& -{1 \over 12} \sum_{k=0}^{N-1} (y_{k+1}-y_{k})(x_{k}^2 + x_{k+1}^2) (x_{k}+x_{k+1})\\
\mathrm{I\textsc{yy}}_{ij}\mathrm A_{ij}&=& +{1 \over 12} \sum_{k=0}^{N-1} (x_{k+1}-x_{k})(y_{k}^2 + y_{k+1}^2) (y_{k}+y_{k+1})\\
\mathrm{I\textsc{xy}}_{ij}\mathrm A_{ij}&=& +{1 \over 6} \sum_{k=0}^{N-1} (x_{k+1}-x_{k}) x_{k} (y_{k}^2 + y_{k+1}^2 + y_{k}y_{k+1}) \nonumber\\
&&+ {1 \over 24} \sum_{k=0}^{N-1} (x_{k+1}-x_{k})^2 (y_k^2+3y_{k+1}^2+2y_{k}y_{k+1})\\
\mathrm{I\textsc{x}}_{ij}\mathrm A_{ij}&=& -{1 \over 6} \sum_{k=0}^{N-1} (y_{k+1}-y_{k})(x_{k}^2+x_{k+1}^2+x_k x_{k+1})\\
\mathrm{I\textsc{y}}_{ij}\mathrm A_{ij}&=& +{1 \over 6} \sum_{k=0}^{N-1} (x_{k+1}-x_{k})(y_{k}^2+y_{k+1}^2+y_k y_{k+1}).
\end{eqnarray}

The sign of Eq.~\ref{eq:aij} corresponds to a specific choice of chirality for the polygonal vertex list. The quantities above are used to evaluate distortions to pixel area ($\delta \ln \mathrm A_{ij}$), pixel astrometric shift vectors ({\it e.g.}, $\vec{p}_{ij}\cdot\hat{x}=[\mathrm{I}\textsc{x}_{ij}\mathrm{A}_{ij}]/\mathrm{A}_{ij}$) and pixel ellipticities ($\delta \epsilon_{1,ij}=[\mathrm{I}\textsc{xx}_{ij}\mathrm{A}_{ij} - \mathrm{I}\textsc{yy}_{ij}\mathrm{A}_{ij}]/[\mathrm{I}\textsc{xx}_{ij}\mathrm{A}_{ij} + \mathrm{I}\textsc{yy}_{ij}\mathrm{A}_{ij}]$; $\delta \epsilon_{2,ij}=2\,\mathrm{I}\textsc{xy}_{ij}\mathrm{A}_{ij}/[\mathrm{I}\textsc{xx}_{ij}\mathrm{A}_{ij} + \mathrm{I}\textsc{yy}_{ij}\mathrm{A}_{ij}]$). Examples for calculations of pixels in the {\it template} are shown in Figure~\ref{fig:template_constraints_geopars} and those evaluated at the end of an exposure for a bright, stellar source for three specific operating conditions are given in Figure~\ref{fig:bf_recordedpars}. 
\end{enumerate}

This internally consistent partition model may have wide applicability to detailed modeling, starting with validation against available laboratory characterization data: flux dependence of flat fields and the shift-covariance statistics they express, modulation transfer function (MTF) measurements, and spot projector illumination using well characterized beams. Proper validation should include quantitative comparisons of the flux dependences for the above tests. In our opinion, the most fundamental property and largest signal to reproduce is the shape of the mean-variance relation obtained from photon transfer measurements. Once validated, the same modeling platform may be used to compare and quantify differences between incident flux distributions and recorded conversion distributions.

\section{Examples}\label{sec:d}
The remainder of this paper is devoted to depicting some results made possible by the updated modeling capabilities described above. In the future, we plan to identify better verification metrics against sensor characterization data acquired under a fixed set of operating conditions - low light levels and voltage settings. Spot projector data~\cite{Tyson_spie_2014} is also of extreme interest, even if the incident beam is not well understood: we are equipped to model and compare changes in the distribution of recorded images as a function of integration level.

\subsection{Tuning dipole strength ratio for consistency with observed covariance term ratio}\label{sec:e}

To tune the dipole strength ratio, we started with the assumption that the bloom stop implant $\vec\xi_{BS}$ is identical to each of the channel stop implants (modulo its symmetry axis $\hat{s}$ orientation). We matched the low-flux flat field response signature to the computed ``fixed pattern'' pixel areas there. Using a backdrop field profile $\vec{E}_{BD}(z)$ corresponding to the $V_{BSS}=-50\mathrm{V}$ setting for that data, we converged on placeholder for the implant distributed dipole moment $\vec\xi_{CS}=-14.3\xi_0\hat{k}$\footnote{This dipole strength is approximately twice as large as that determined previously~\cite[Fig.5]{Rasmussen_spie_2014}. This difference is partially explained by the change in the symmetry of the electrostatic problem, where previously two equipotential planes were used, and a truncated infinite expansion of image charges were used.}, where as previously, $\xi_0\equiv 10^{-6}\mathrm{q_e}=1.6\times 10^{-25}\,\mathrm{Coul}$. This choice also provided edge rolloff pixel area profiles that matched reasonably well when compared against the flat field response signature there. Considering the value of $\vec\xi_{CS}$ to be fixed, the value of the clock rail induced dipole moment $\vec\xi_{CK}$ was constrained by computing the ratio of geometric area variations for neighboring pixels due to presence of accumulated charge in the ``aggressor'' pixel\footnote{For convenience and for an initial tuning this model, we adjust the value of $\vec\xi_{CK}$. What is really being adjusted in so doing is the position within the photosensitive bulk where the backdrop field $\vec{E}_{BD}(z)$ is nearly canceled by the collective $\delta \vec{E}_\perp^{CK}$. There are multiple solutions that could provide this knob, including a modification of the assumed doping profile through the photosensitive bulk that provide the built-in field~\cite{Rasmussen_paccd_2014,Rasmussen_spie_2014}. We have not identified other independent observables that constrain the built-in field, so for now we choose to leave $\vec{E}_{BD}(z)$ fixed. In the future, the dependence of covariance terms on BSS may be used to jointly constrain $\vec{E}_{BD}(z)$ and $\delta \vec{E}_\perp^{CK}$.}. This ratio is plotted in Figure~\ref{fig:template_constraints_geopars} (upper left panel) for a range of backside bias voltages and for a selection of trial clock dipole values. We identified a ``fiducial performance'' region of this parameter space that exhibits the observed ratios in geometric area variations, inferred directly from covariance ratios~\cite{Antilogus_paccd_2014}. The ``tuned'' value for the clock dipole magnitude was then chosen to be $|\vec{\xi}_{CK}|=5\,\xi_0$. This specific combination of contributors to $\vec{E}^{tot}$ now features pixel neighbor area perturbations that respond to collected charge (the brighter-fatter effect) with proper proportions. It should be noted that a joint fit could also have been performed to match the various observables. For example, such fits may favor a different dipole strength $\vec\xi_{BS}$ for the bloom stop barrier, a different acceptor density $N_a$ for the photosensitive bulk, or a different field strength offset when $V_{BSS}=0$ and the device is fully depleted. We have not yet performed such fits that may further tune parameters of this model.

The influence of collected charge at the channel remains to be calibrated. This is a straightforward step that is best done using constraints of the mean-variance relation. It is equivalent to determining the depth of the buried channel in this simplistic model, because the dipole moments $\vec{p}_{ij}$ scale with that dimension. We also expect that the depth or position of the potential well will vary with the number of conversions collected. These details will depend on a number of factors, including the doping profile of the n-type channel and the attractive influence of the image charge that, combined together with the collected charge cloud in the channel and the presence of adjacent, similar, conversion-induced dipole moments in neighboring pixels, self-consistently form the dipole moments $\vec{p}_{ij}$. We assume that such details may be most easily accessed from photon transfer data sets. For now, we consider only potential well depths that are constant with integrated conversions.

The resulting brighter-fatter {\it template} was generated for $\vec{p}_{ij}=-1.0\,\delta_{i0}\delta_{ij}\mathrm{p}_0\hat{k}$ (where $\mathrm{p}_0\equiv 10^5\,\mathrm{q_e}\mu\mathrm{m}$) and is available in the form of pixel specific boundary contours (Figure~\ref{fig:bf_recordedpars}, upper left panel) and the pixel geometric moments obtained using the shoelace formulae given above. The latter are plotted in Figure~\ref{fig:template_constraints_geopars}, upper right and lower panels. The overall radial dependence of these parameters (lower panel) provides an overview of the brighter-fatter effect, with radial dependences evaluated by fitting over the range $4 \le (\Delta i)^2+(\Delta j)^2\le 25\,\mathrm{pix}^2$.
The following radial scaling dependences were obtained for the fitting range: $\Delta \delta \ln A_{ij} / \Delta \ln r^2 \approx -1.288$, $\Delta \delta \ln (1-p_{r,ij}) / \Delta \ln r^2 \approx -1.033$ and $\Delta \delta \ln (1-\epsilon_{r,ij}) / \Delta \ln r^2 \approx -1.410$. Here, the radial component of ellipticity for pixel $(i,j)$ is defined as $\epsilon_{r,ij} \equiv (I_{rr}-I_{\theta\theta})/(I_{xx}+I_{yy})$. Notice that these slopes are dependent on the fit domain applied. Two features should be pointed out: First, the geometric distortions for distances within the fit domain and farther - should be relatively insensitive to the influence of $\vec\xi_{CS}$ and $\vec\xi_{CK}$, because most of the boundary shifts occur as a result of curved drift lines where the backdrop field $\vec{E}_{BD}$ is smallest. This also implies that this component of the geometric distortions should have some chromatic dependence. The geometric distortions appear to fall off rapidly beyond 10~pixels -- which happens to be equal to the assumed thickness of the sensor. This is reassuring, because influences fall off more quickly than $r^{-2}$, and a limited number corrections might be required in any pixel pipeline that utilizes them. Second, the large deviations from the power law fits for the nearest pixels are unique and the short distance effect is clearly not isotropic (a consequence of the tuned dipole moment ratios). The drift contribution to these geometric distortions must be small, because the lateral electrostatic influence of a dipole field for points along the dipole's axis vanishes. What appears to govern the pixel boundaries closest to the collected conversions, are the distortions of the saddle point loci experienced by charges flowing along the drift lines close to the pixel boundaries. This is the component of pixel geometric distortion that can't be simulated with simplified electrostatic drift models. Because these shifts in pixel boundaries occur close to the gates and channel stops, this contribution should be nearly achromatic.

\subsection{Large amplitude fixed pattern features}\label{sec:f}

\begin{figure}[tbp]
\begin{tabular}{rr}
\includegraphics[trim=1.5cm 1.5cm 2cm 1.5cm, clip=true, width=.45\textwidth]{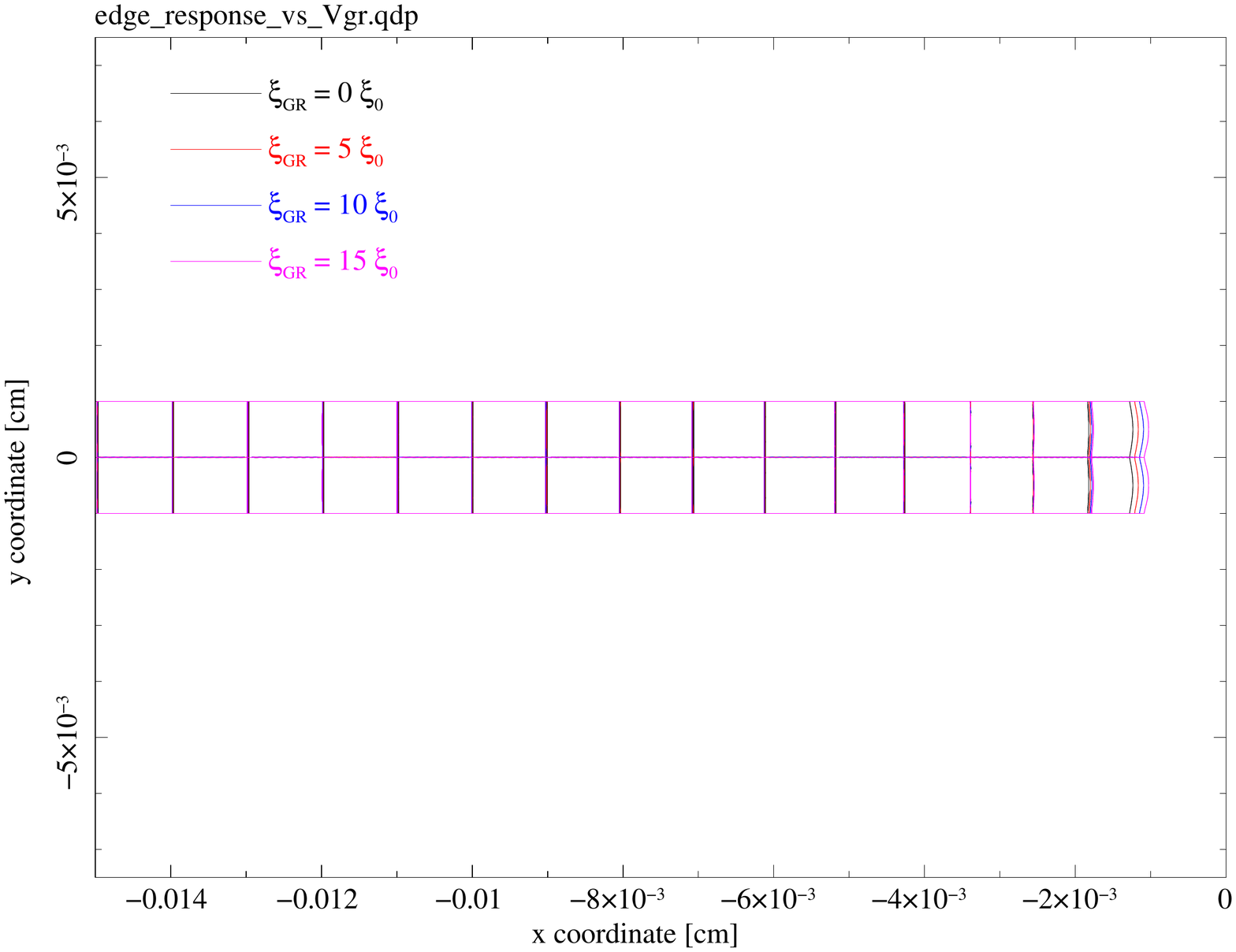}&
\includegraphics[trim=1.5cm 1.5cm 2cm 1.5cm, clip=true, width=.45\textwidth]{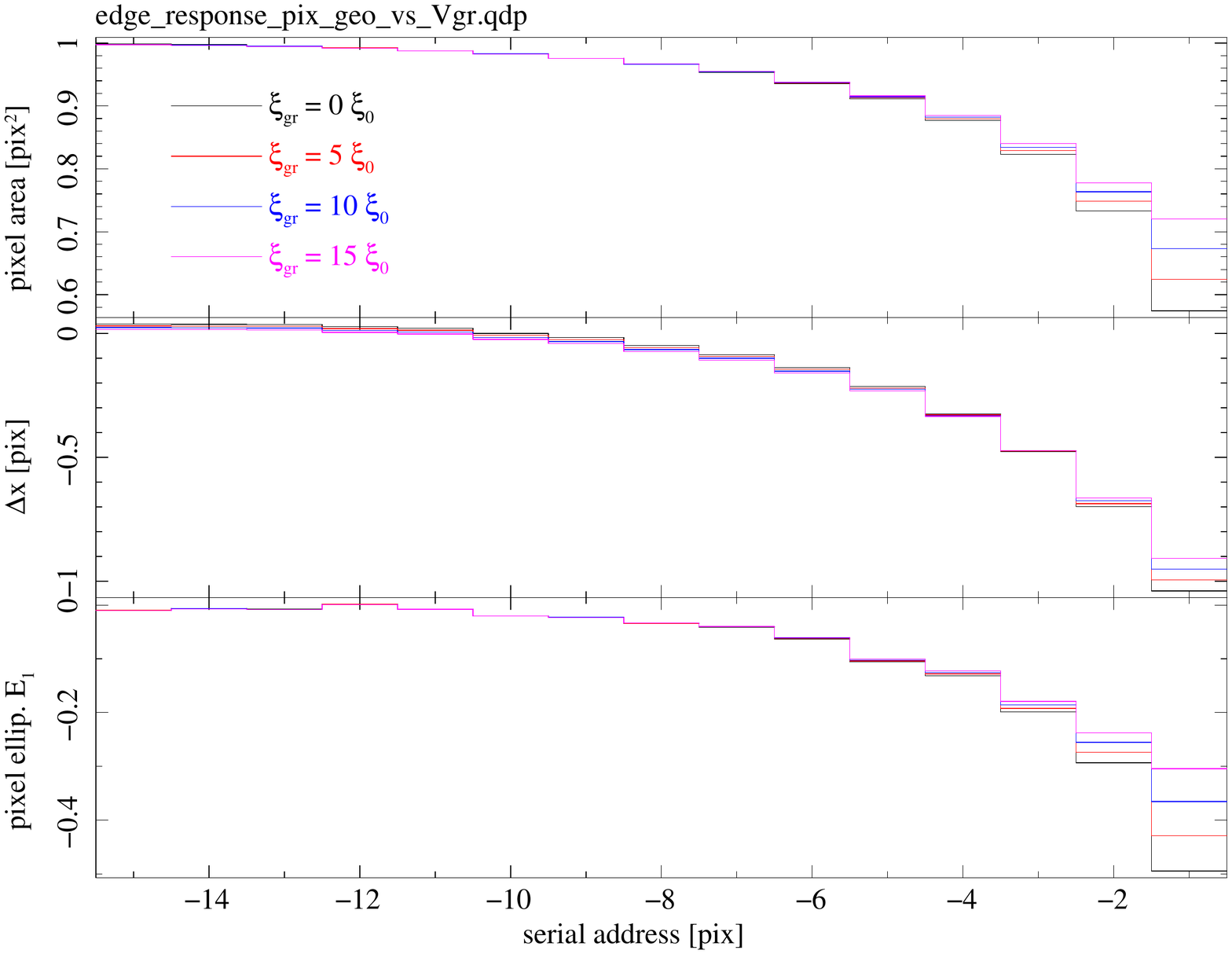}\\
\includegraphics[trim=1.5cm 1.5cm 2cm 1.5cm, clip=true, width=.45\textwidth]{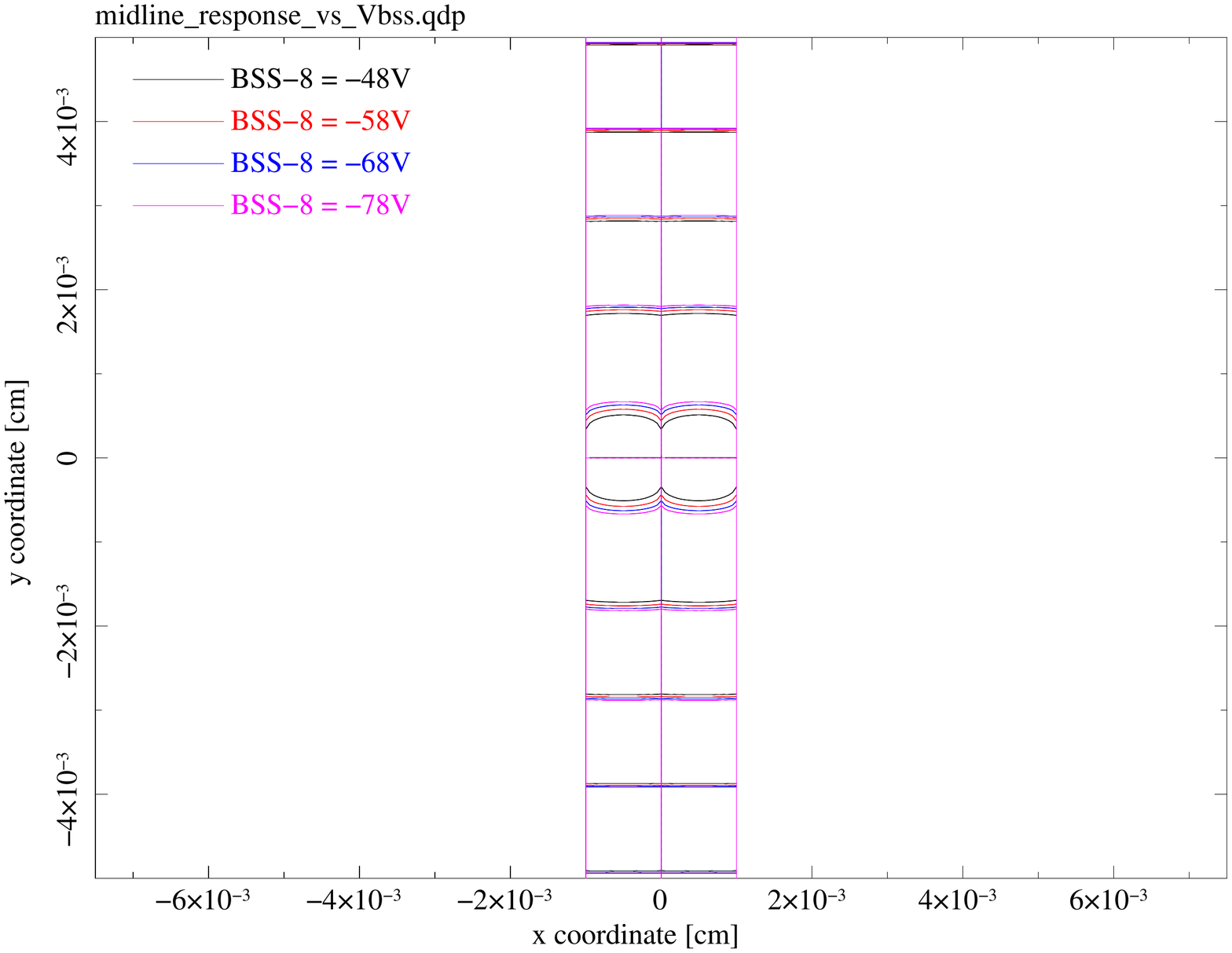}&
\includegraphics[trim=1.5cm 1.5cm 2cm 1.5cm, clip=true, width=.45\textwidth]{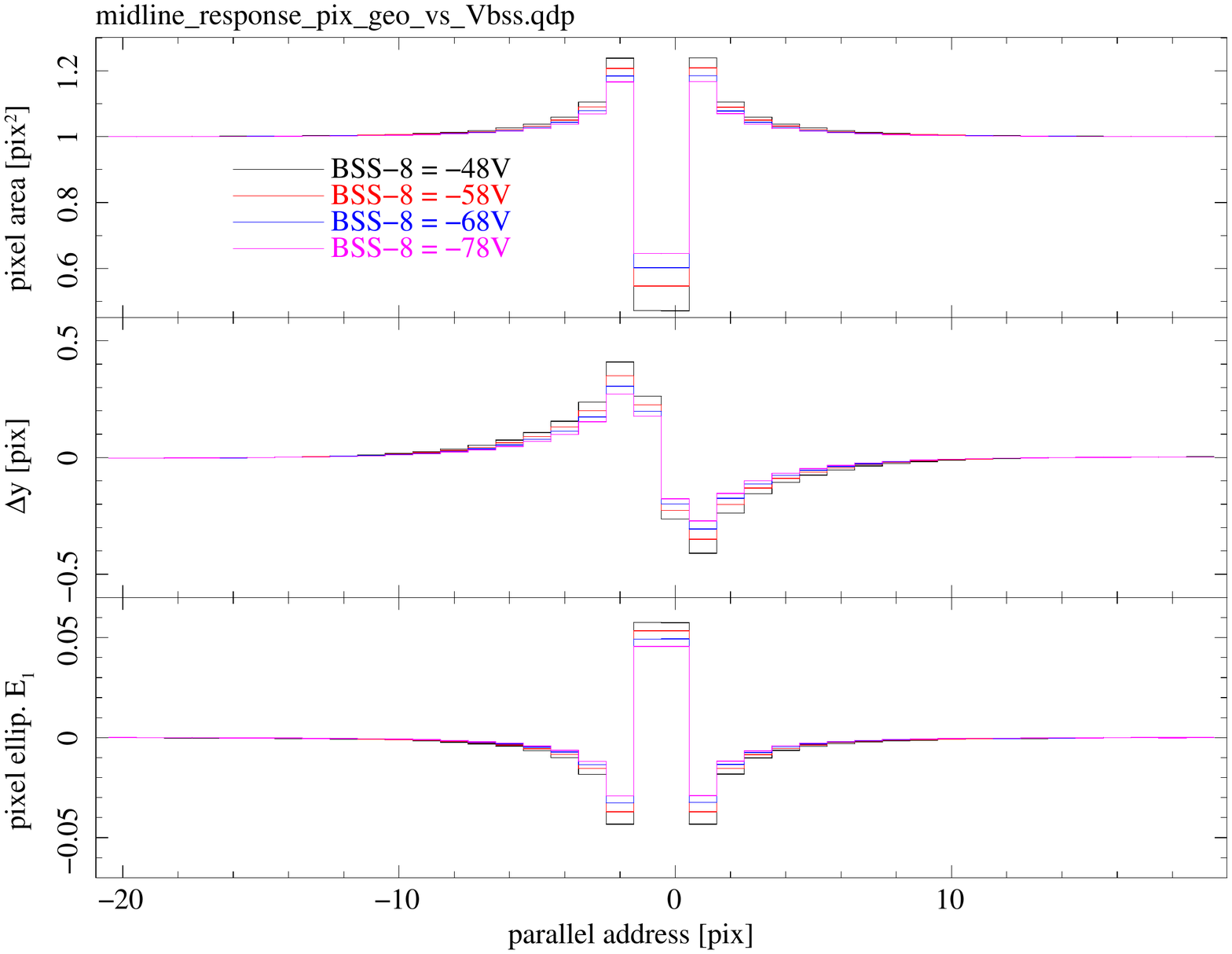}\\
\end{tabular}
\caption{Edge rolloff response (top plots) and midline redistribution (bottom plots) large amplitude fixed pattern feature pixel calculations. For each effect, pixel boundary contours are shown for two rows or columns (left hand plots), while geometric parameters $A_{ij}$, $\vec{p}_{ij}$ and $\vec\epsilon_{ij}$ are shown (right hand plots). Predictions are given for adjacent guard ring drain bias settings (top) and for backside bias settings (bottom).}
\label{fig:edge-n-midline}
\end{figure}

\begin{figure}[tbph]
\begin{tabular}{cc}
\includegraphics[trim=1.5cm 1.5cm 1.5cm 1.5cm, clip=true, height=0.4\textwidth]{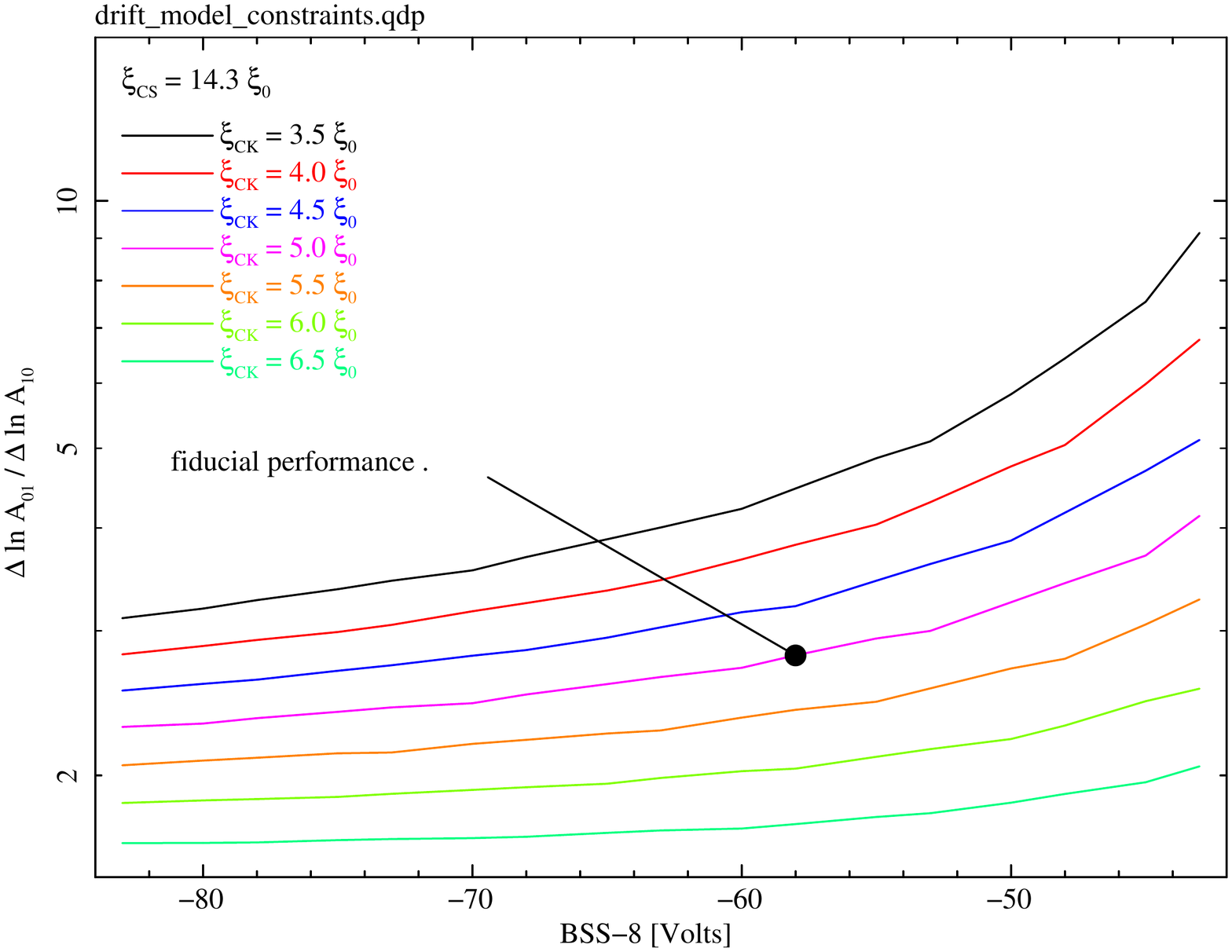}&
\includegraphics[trim=1.5cm 0cm 0.5cm 0cm, clip=true, height=0.4\textwidth]{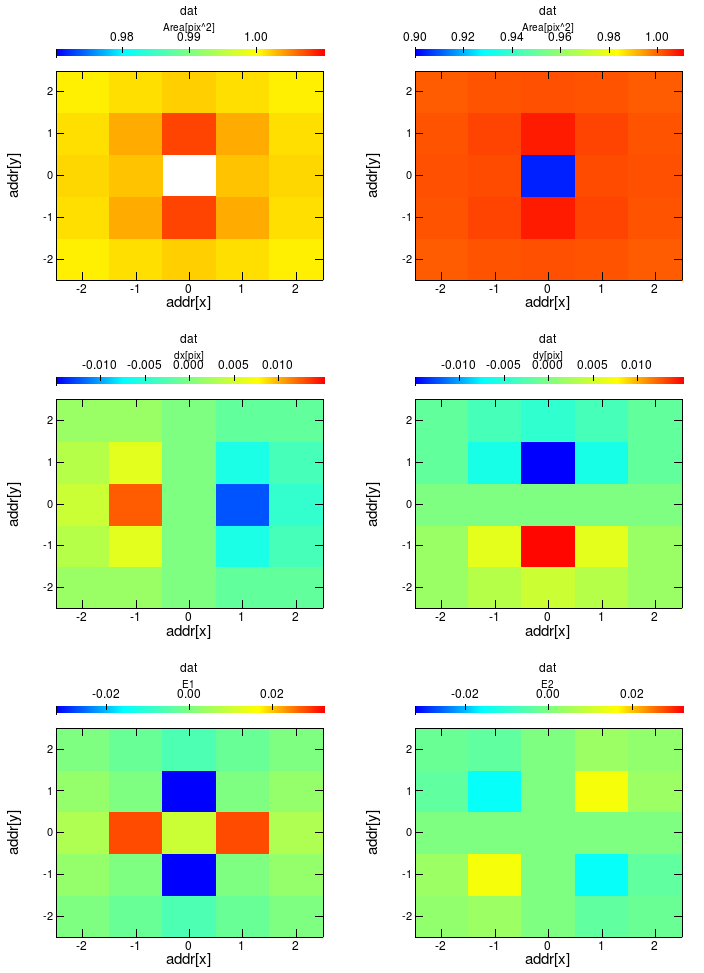}\\
\multicolumn{2}{c}{\includegraphics[width=0.9\textwidth]{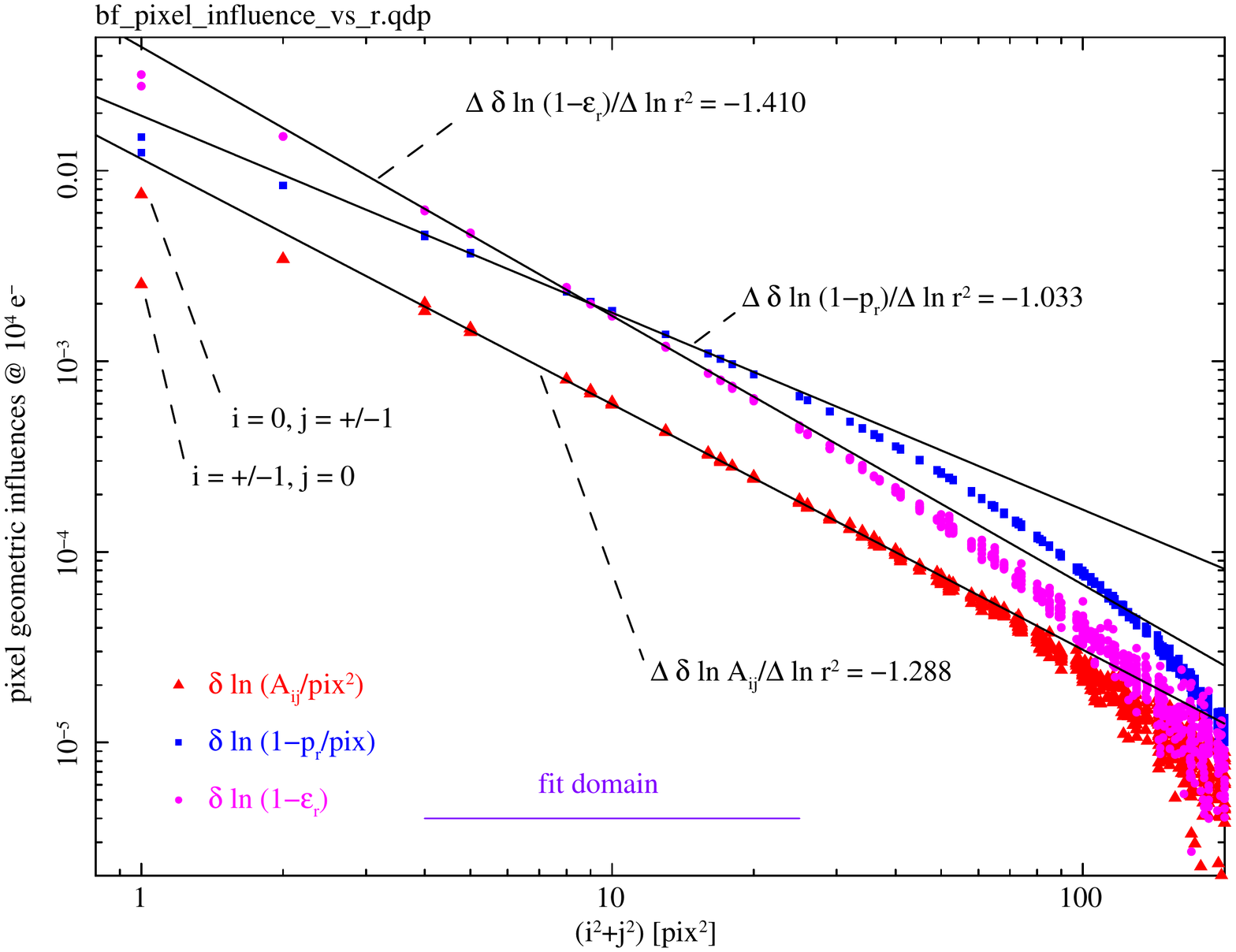}}
\end{tabular}
\caption{Properties of the brighter-fatter pixel distortion template. Upper left: the dependence 
of the area distortion ratio $\delta\ln A_{01}/\delta\ln A_{10}$ on the backside bias BSS for a 
range of clock induced dipole moment values $\xi_{CK}$. The round black dot indicates the 
fiducial performance (parameter choice) for the remainder of this paper. Upper right: a false 
color representation of the pixel geometric parameters in the template, arranged by address: 
areas $A_{ij}$ (top two), pixel astrometric shifts $\vec{p}_{ij}$ (middle two), and pixel ellipticity 
$\vec\epsilon_{ij}$ (bottom two). Bottom: pixel geometric parameters, this time plotted against 
separation ($r^2=i^2+j^2$) from the central pixel. Of these parameters, only $\delta\ln A_{ij}$ 
are constrained directly by high signal-to-noise, exposure corrected flat field images. 
The radial 
dependence agrees within a few percent of the dependence found for a completely different device (DECam) where a full season's worth of $r$-band dome flats were jointly used for 59 devices (\cite[Fig.4]{Gruen_paccd_2015}) -- but this drift model also explains the nature of the high significance, high amplitude outliers seen there. The assumed 
dipole of the aggressor for this template was $\vec{p}_{00}=-\mathrm{p}_0\hat{k}$. See text for more 
information.}
\label{fig:template_constraints_geopars}
\end{figure}

We evaluate the pixel boundaries near the edges (parallel to the parallel address axis) and those crossing the midline bloom stop. Using the fiducial performance parameters outlined above, we obtain the pixel contours, pixel areas $A_{ij}$, pixel position shifts $\vec{p}_{ij}$ and pixel ellipticities $\vec{e}_{ij}$ for the affected pixels. They are summarized in Figure~\ref{fig:edge-n-midline}, where we have overlaid those quantities for different guard ring drain bias induced dipole moments $\vec\xi_{GR}$ (for the edge response case). We have also overlaid quantities for different backside bias values (for the midline redistribution case). In all cases we predict subtle variations in the detailed pixel shapes when evaluated at the backside surface - the edges often appear to be scalloped in shape and the notion of parallel pixel boundaries is generally violated.

\subsection{Flat field simulations that include dynamic pixel boundary distortions}\label{sec:g}

Here we simulate flat fields to predict the relevant statistical properties of those illuminations. Recall that in \S\ref{sec:e} above, the clock dipole strength $\vec\xi_{CK}$ was adjusted so that the ratio of pixel area perturbations would match the measured pixel covariance ratio: $\delta \ln A_{10}/\delta \ln A_{10} = C_{10}/C_{01} \sim 3$. This is a check to test the validity of that assumption. Starting with the brighter-fatter template described in Figure~\ref{fig:template_constraints_geopars}, we populate a stack of 1024 small images with a randomly distributed, constant flux to represent a flat field illumination. This is done in steps of on average 1000 conversions per pixel, using a Poisson deviate together with the pixels' current areas $A_{ij}$. Between population steps, some statistical calculations are performed on the recorded charge patterns in the image stack, and those patterns are used to reevaluate the pixel areas that feed into the next population cycle. The statistical calculations are performed much in the same way as laboratory data would be: for every unique pair of images, a difference image is computed. For specific serial and parallel shift (lag) values, product of the difference values is summed to contribute to an overall covariance. The grand total sum is divided by the total number of electrons contributing to the calculation ($2\mu\mathrm{N}_\mathrm{pix}$) to yield the covariance value $C_{\Delta i,\Delta j}$ for lag ($\Delta i,\Delta j$). When performed for a single image pair, the covariance is simply the numerator in the expression for a difference image's autocorrelation element.

The calculation is complete after 250 such population and analysis cycles. Two assumed responsivities were applied to the mechanism. The first assumes that the brighter-fatter template corresponds to an isolated channel population of $10^4$ conversions, at which point the central pixel area shrinks to $A_{00}=0.9033$; adjacent pixel areas grow modestly to become $A_{01}=1.0077$, $A_{10}=1.0027$ and $A_{11}=1.0033$ ($\delta \ln A_{00}=-1\times 10^{-5}/\mathrm{e}^-$). The second assumed that the influence is 3.3 times as strong, for the same channel population ($\delta \ln A_{00}=-3.3\times 10^{-5}/\mathrm{e}^{-}$). When we compare these computed $C_{00}$ curves to carefully prepared photon transfer curves (\cite[Fig.4]{Guyonnet_2015}), it is clear that a more realistic responsivity of this effect would be about a factor of 3 smaller than the first case, or $\delta \ln A_{00}=-3\times 10^{-6}/\mathrm{e}^{-}$.

\begin{figure}[tbp]
\begin{tabular}{rr}
\includegraphics[trim=1.5cm 1.5cm 1.5cm 1.5cm, clip=true, width=.5\textwidth]{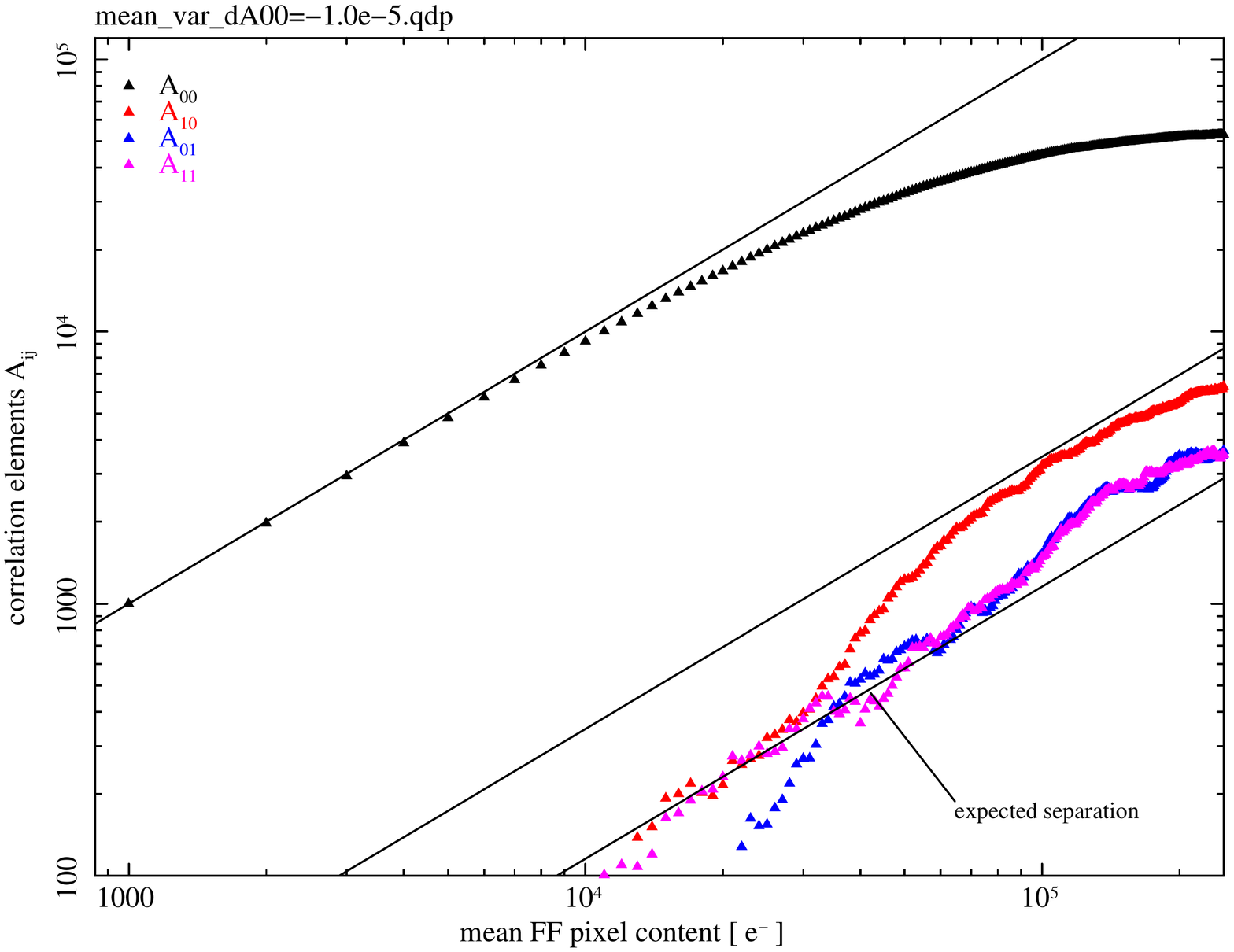}&
\includegraphics[trim=1.5cm 1.5cm 1.5cm 1.5cm, clip=true, width=.5\textwidth]{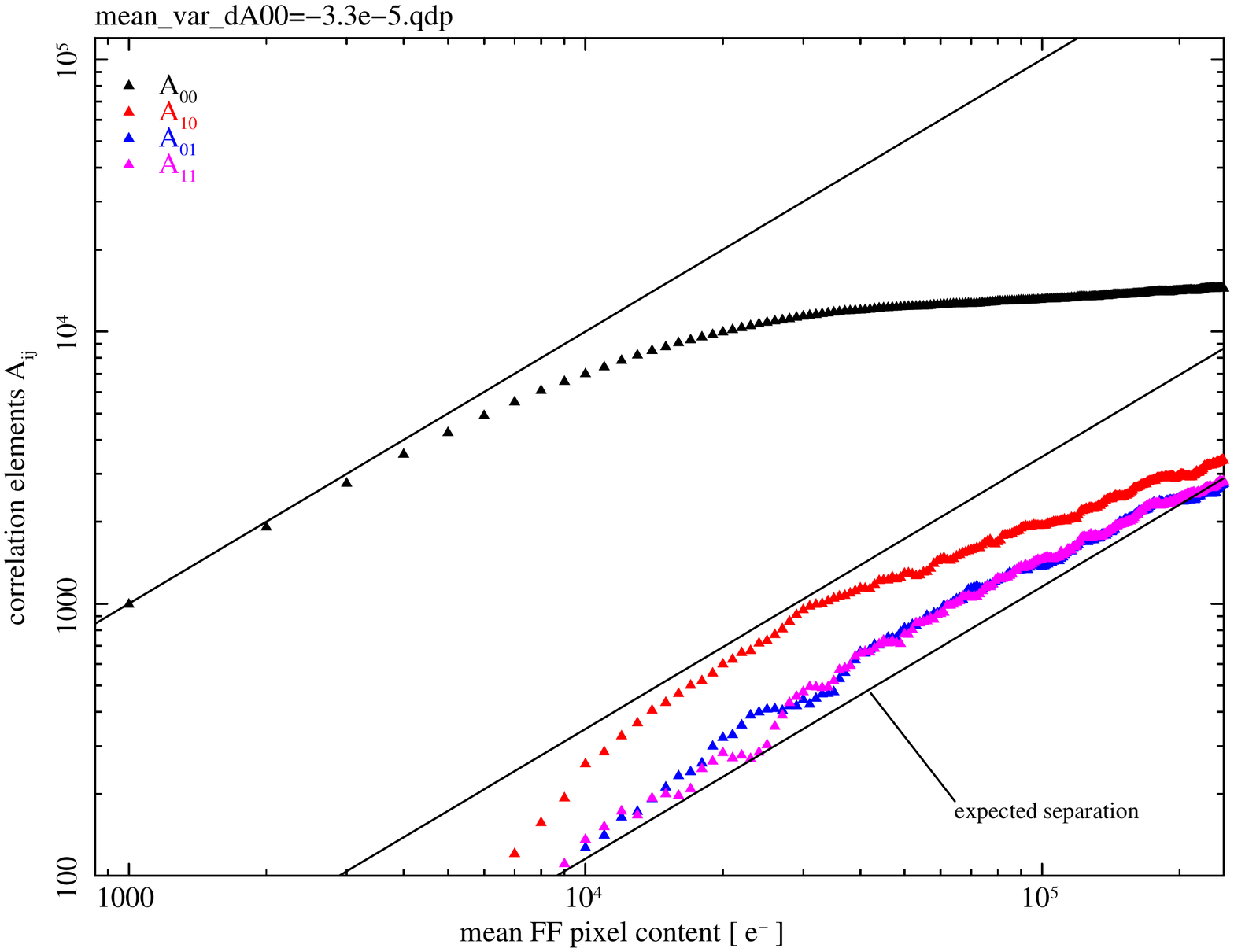}\\
\end{tabular}
\caption{Covariances $C_{ij}$ of difference images where a lag ($i,j$) has been applied in the calculation. The covariance $C_{00}$ is simply the variance. Other terms $C_{01}$,  $C_{10}$ and $C_{11}$ are numerators in the autocorrelation of the difference images. These terms rise as $\mu^2$ for low light levels, where also the variance deviates only linearly from the Poisson behavior linearly with $\mu$. At intermediate values of $\mu$, where the {\it deviations} from Poisson behavior become dominant, $C_{ij}$ grow linearly with $\mu$, and then flatten off at larger values. The straight lines showed here provide guides for Poisson behavior in $C_{00}$, and for the linear growth region for $C_{01}$ and $C_{10}$, based on area perturbations $\delta\ln A_{01}$ and $\delta\ln A_{10}$ computed from the brighter-fatter template. Two different response levels for the brighter-fatter effect were assumed: $\delta\ln A_{00}=-1\times 10^{-5}/\mathrm{e}^-$ (left) and $-3.3\times 10^{-5}/\mathrm{e}^-$(right). Carefully measured curvature of the mean-variance curve for specific operating conditions~\cite[Fig.4]{Guyonnet_2015} would calibrate the response to $\sim -3\times 10^{-6}/\mathrm{e}^-$. Pixel areas were reevaluated for every 1000 electron increment to the flat field level, or once per plotting symbol shown.}
\label{fig:mean-var}
\end{figure}

Figure~\ref{fig:mean-var} provides the results of these calculations, where we plot only the mean-variance relation $C_{00}$ and the three covariance terms $C_{01}$, $C_{10}$ and $C_{11}$. One plot is given for each responsivity assumed. Notice that the transfer of variance from the central pixel containing statistical fluctuations, out to adjoining pixels whose areas adjust in response, is complex enough that pixel area distortion ratios may not be accurately inferred directly from covariance term ratios, at least for the calculation provided here. Potential complications to the working model include at least the following:  1) only a fraction ($\sim$~35\%) of the a given pixel's area perturbation due to its contained charge excess or deficit, is recovered by the 8 that directly adjoin it; 2) that there are multiple aggressors that affect an individual pixel's area perturbation; and 3) this generalized photon transfer calculation exhibits linear behavior only for the lowest light levels, where these calculations clearly fall short statistically. For the near term, we would like to quantitatively understand details of the mean-variance curve's shape ($C_{00}$) in sensor characterization data, since to first order its area perturbation is largely driven by its own charge excess or deficit.

\subsection{Point source response using the brighter-fatter template as a Greens function}\label{sec:i}

\begin{figure}[tbp]
\begin{tabular}{rr}
\includegraphics[trim=1.5cm 1.5cm 1.5cm 1.5cm, clip=true, width=0.48\textwidth]{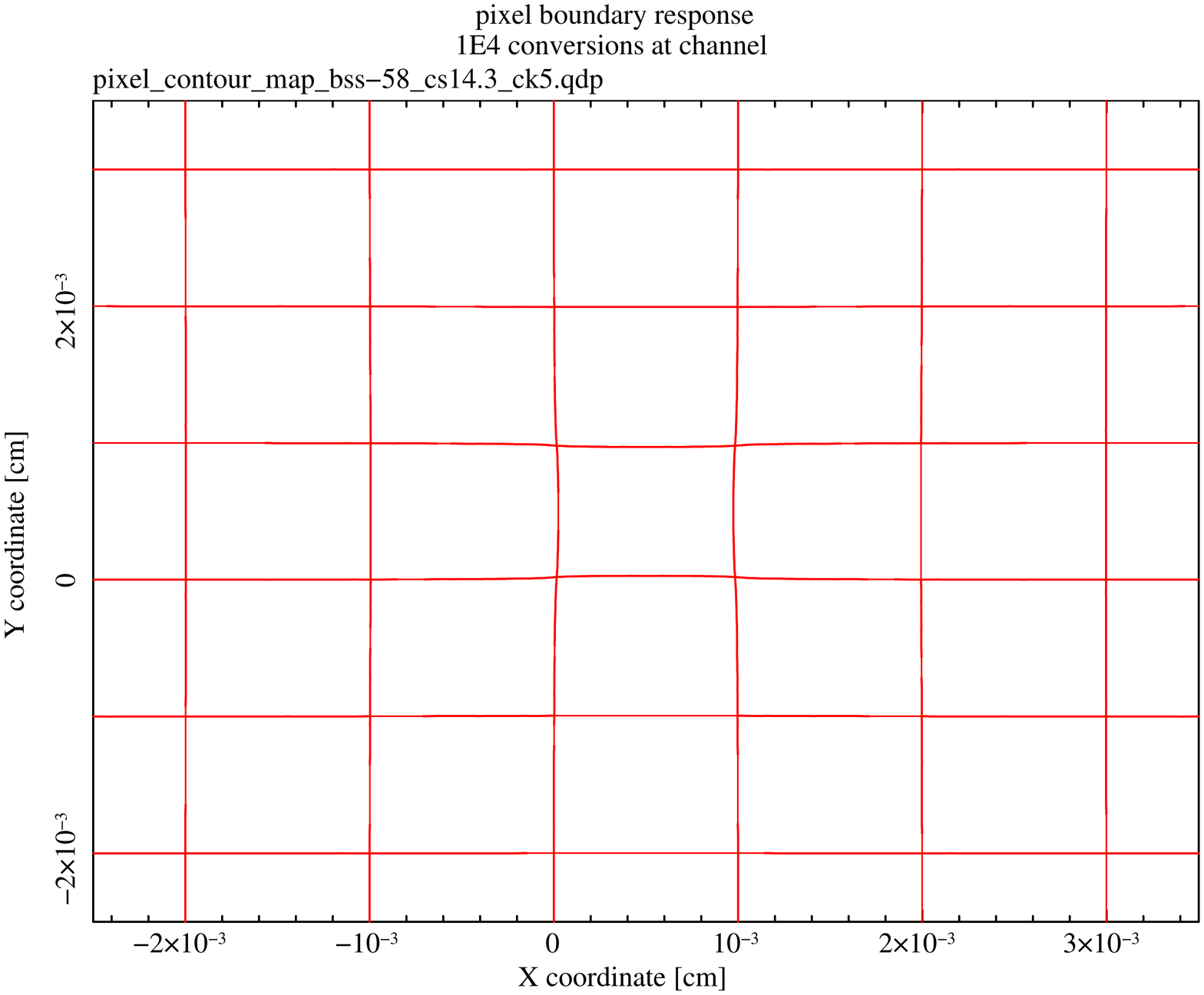}&
\includegraphics[trim=1.5cm 1.5cm 1.5cm 1.5cm, clip=true, width=0.48\textwidth]{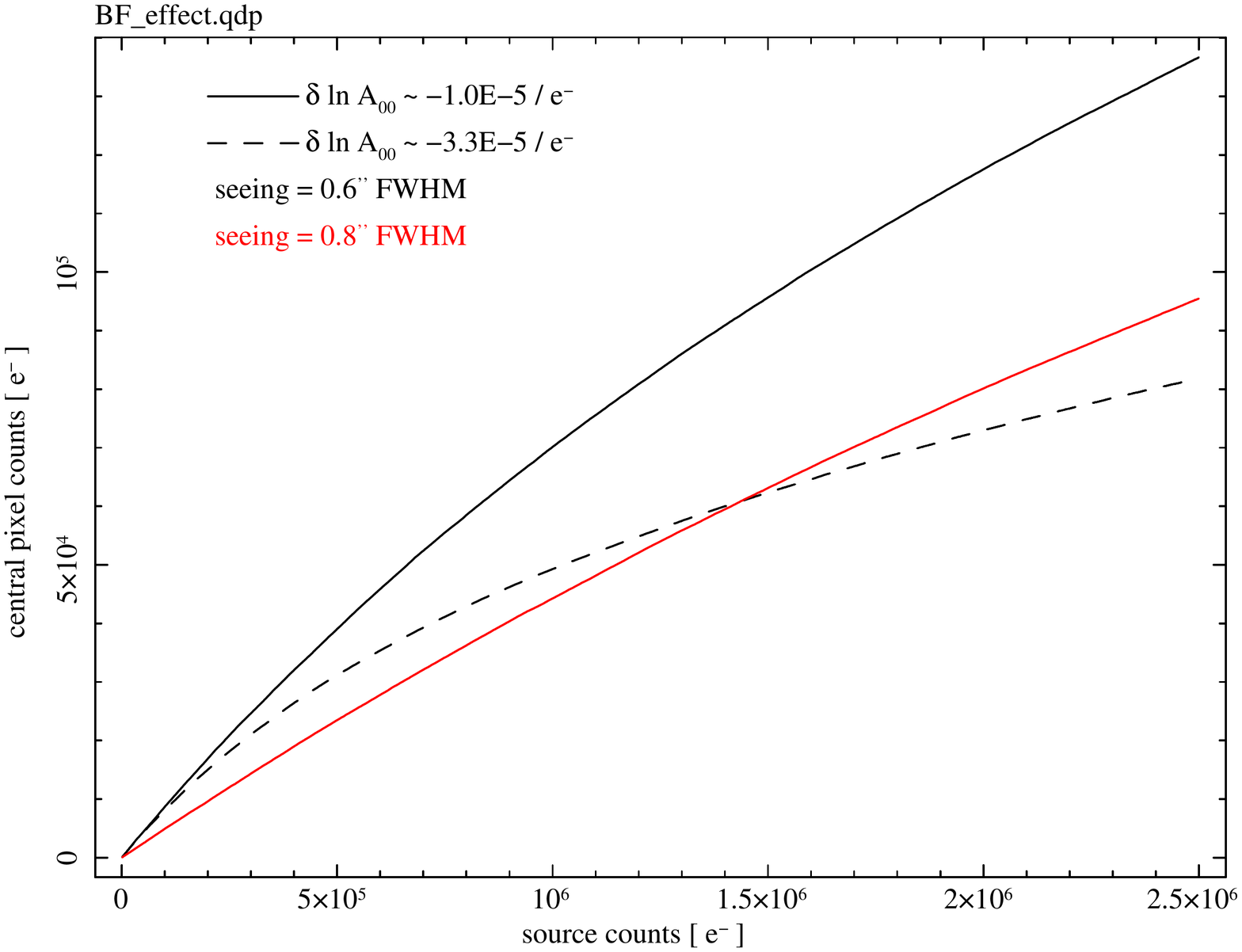}\\
\includegraphics[trim=1.5cm 1.5cm 1.5cm 1.5cm, clip=true, width=0.48\textwidth]{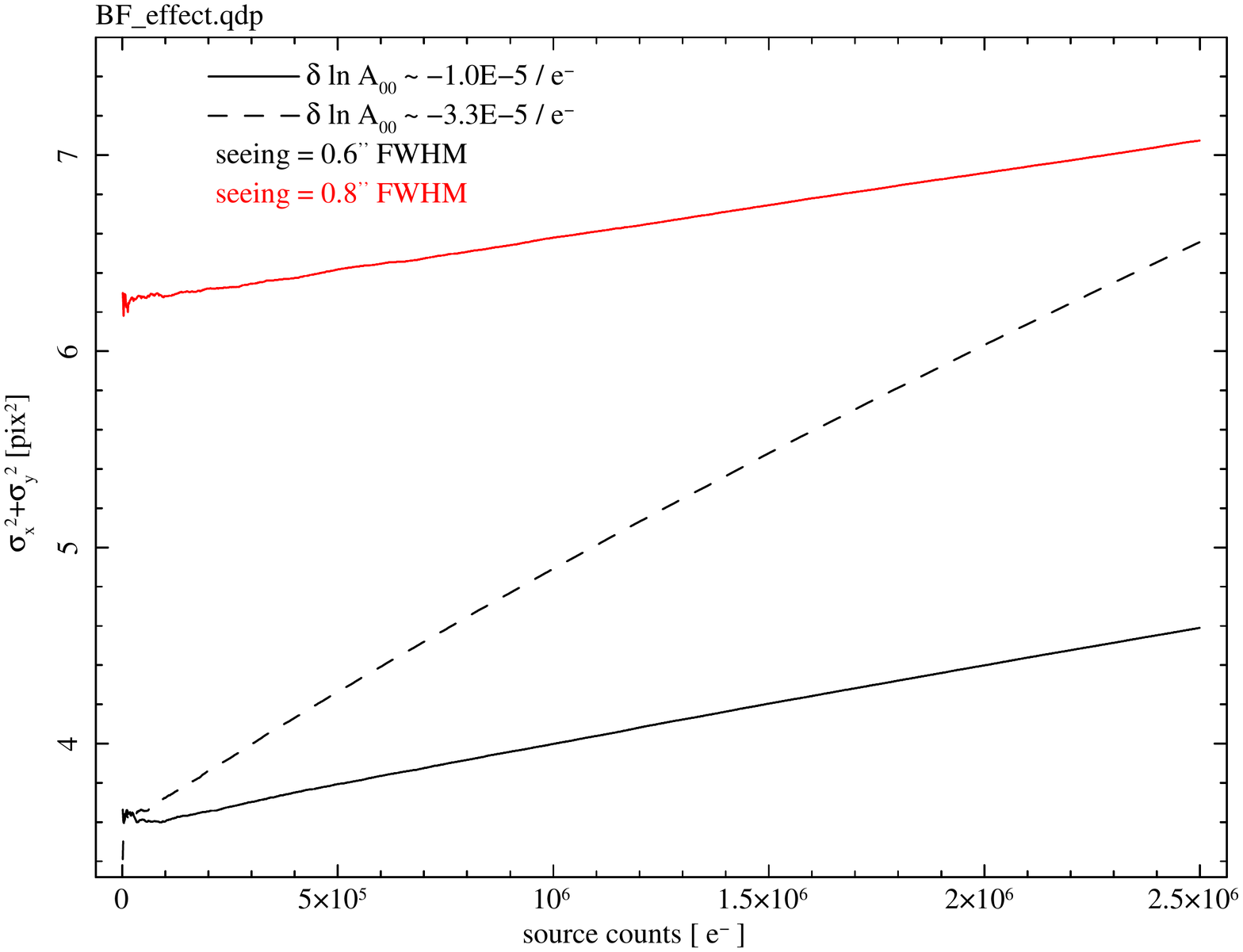}&
\includegraphics[trim=1.5cm 1.5cm 1.5cm 1.5cm, clip=true, width=0.48\textwidth]{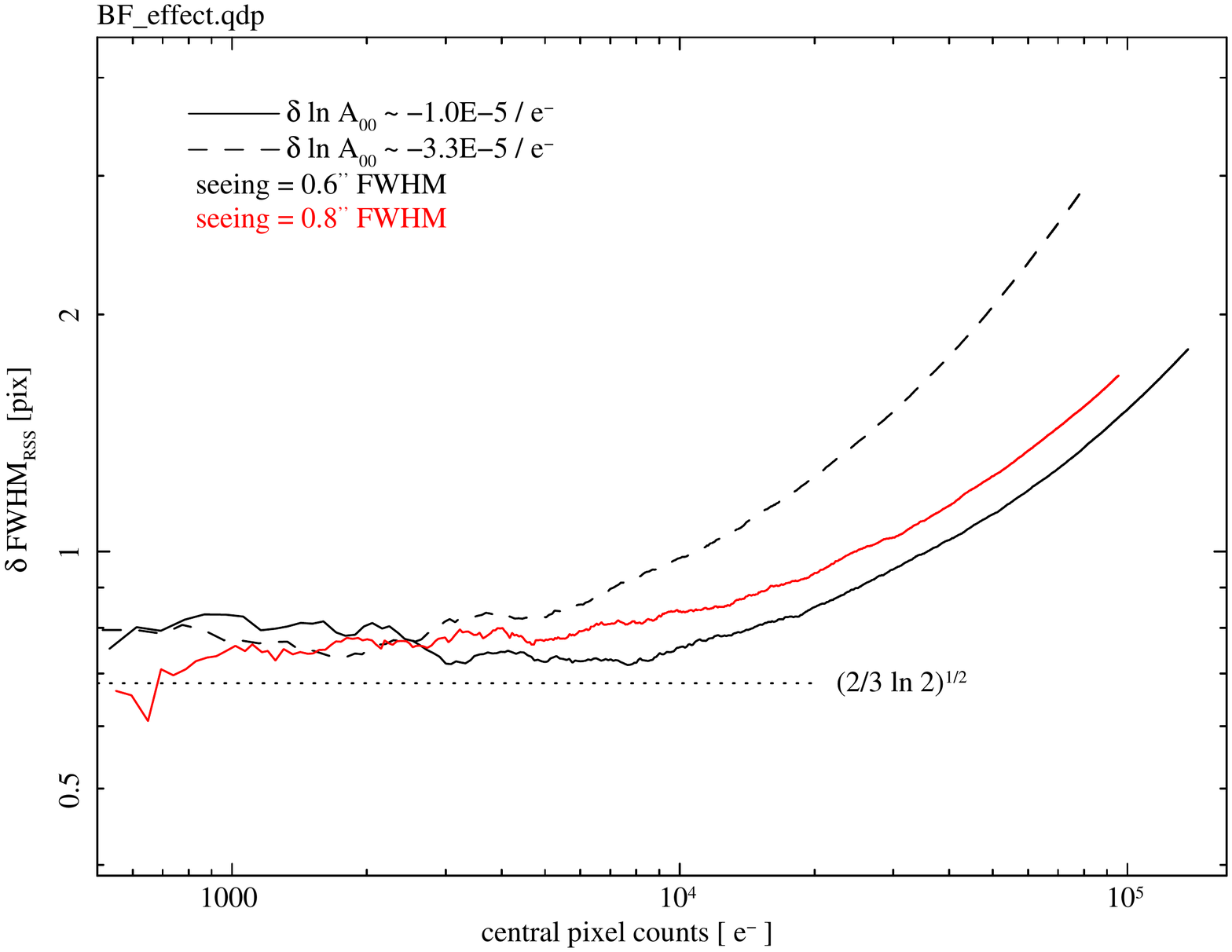}\\
\includegraphics[trim=1.5cm 1.5cm 1.5cm 1.5cm, clip=true, width=0.48\textwidth]{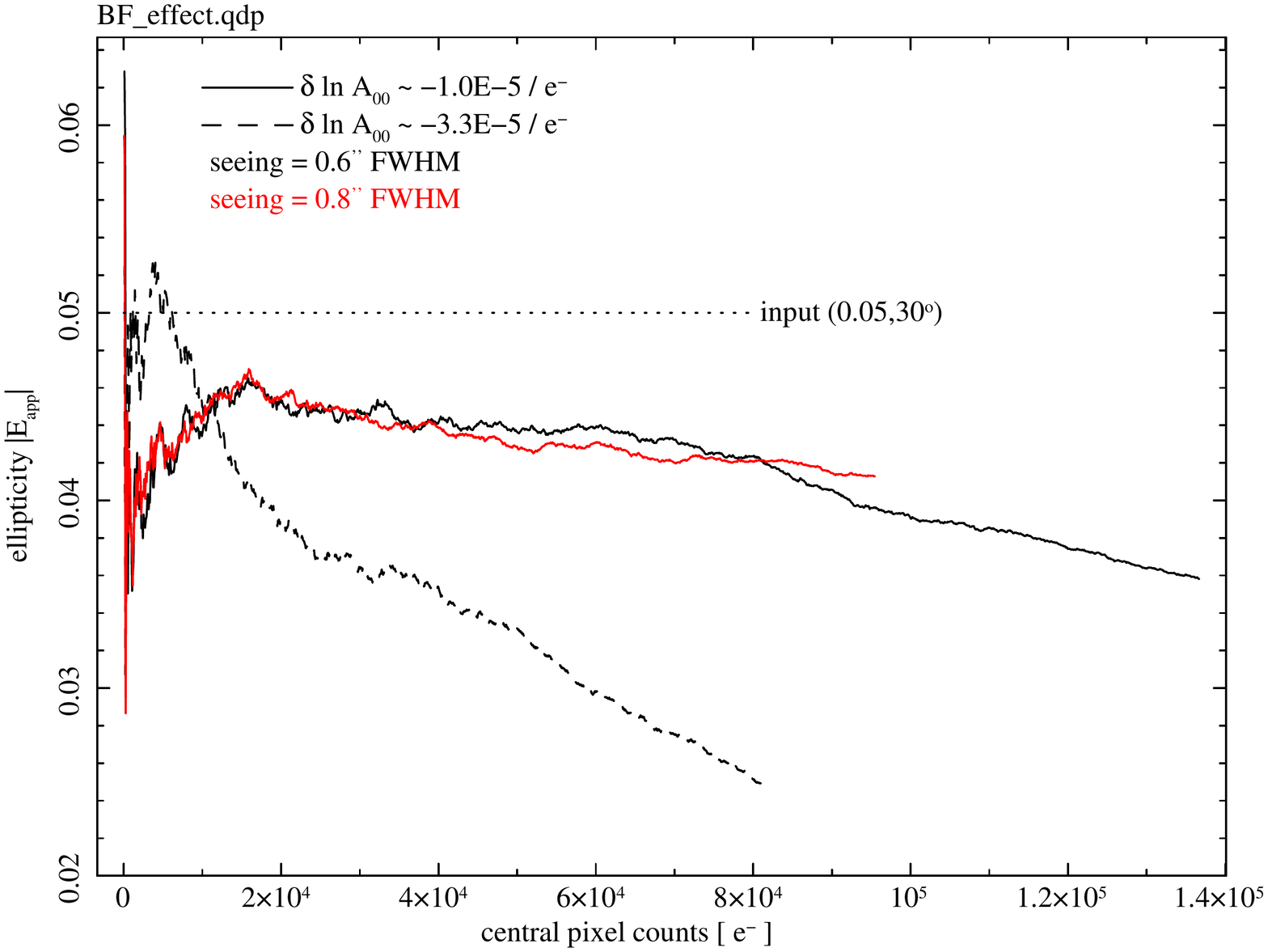}&
\includegraphics[trim=1.5cm 1.5cm 1.5cm 1.5cm, clip=true, width=0.48\textwidth]{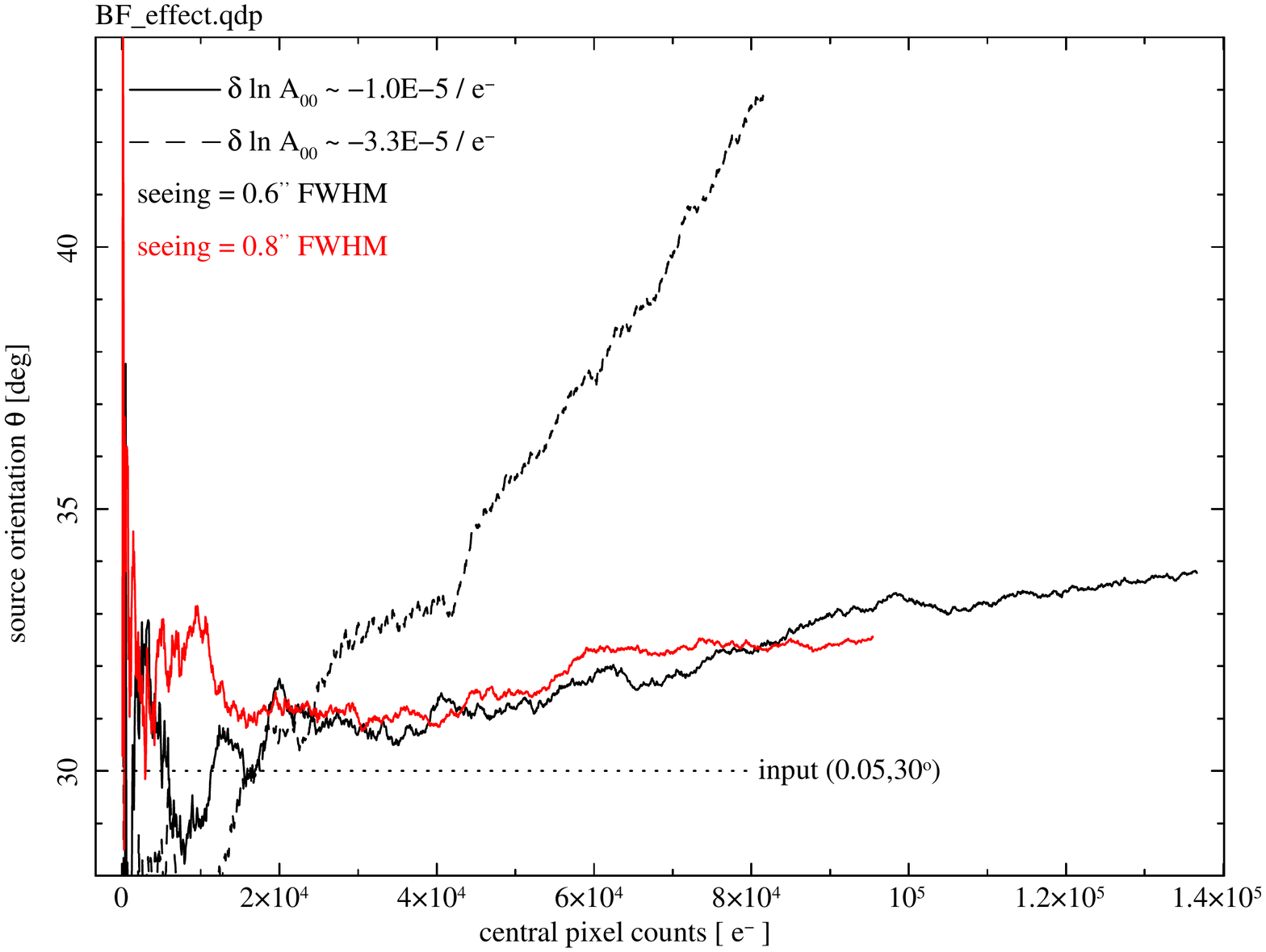}\\
\end{tabular}
\caption{Summary of the three focused image simulations performed for pixel recording properties. Upper left: a display of the pixel boundary contours for the brighter-fatter template used. Upper right: central pixel counts, arranged by PSF counts. Middle left: recorded PSF area $\sigma_i^2+\sigma_j^2$, arranged by source counts. Middle right: FWHM contribution by the brighter-fatter effect (RSS difference) arranged by central pixel counts. Lower left: ellipticity of the recorded image, arranged by central pixel counts. The drop in ellipticity occurs because the input source had a modest ellipticity (0.05) oriented at 30$^\circ$, whereas the brighter-fatter effect is anisotropic. Lower right: recorded source orientation, based on ellipticity components. This shows the anisotropic properties of the brighter-fatter effect, as brighter objects will be systematically rotated toward a 90$^\circ$ orientation.}
\label{fig:bf_recordedpars}
\end{figure}

An initial investigation into how drift calculations can be used to predict errors in PSF estimation, we utilized the same brighter-fatter template in a Monte-Carlo to simulate image formation for an unresolved point source. For the PSF model, we drew sample positions one at a time from an ellipsoidal Gaussian function, with FWHM set by two possible seeing values (0.6$^{\scriptstyle{''}}$ and 0.8$^{\scriptstyle{''}}$) and an ellipticity of 0.05. The major axis is rotated $30^\circ$ counter clockwise from the serial address axis, with centroid positioned at (0.25,0.25) pixel units measured relative to the central pixel's centroid. Starting with an undistorted array of pixel boundaries, sample positions were partitioned into the grid of registering pixels using a point-in-polygon algorithm. Conversions were accumulated 1000 at a time, which typically boosted the central pixel content by about either 80 or 50 conversions, depending on the seeing level in effect. Moments of the accumulating image were computed, as were changes in the image since last update. Following this, the brighter-fatter template was used to update the pixel boundaries based on the current, collected charge pattern. This operation closely resembles what was done for the flat field simulations ({\it cf.} \S\ref{sec:g}), except that conversions were added one at a time.

Results of this image accumulation calculation are given in Figure~\ref{fig:bf_recordedpars}. As in the case of the flat field integration examples above, the same two brighter-fatter responsivities were applied, which correspond to central pixel shrinking rates of $\delta \ln A_{00}=-1.0\times 10^{-5}/\mathrm{e}^{-}$ and $\delta \ln A_{00}=-3.3\times 10^{-5}/\mathrm{e}^{-}$. A total of three calculations are summarized. For $\delta\ln A_{00}=-1.0\times 10^{-5}/\mathrm{e}^-$, seeing levels of 0.6$^{\scriptstyle{''}}$ and 0.8$^{\scriptstyle{''}}$ were simulated, whereas for $\delta\ln A_{00}=-3.3\times 10^{-5}/\mathrm{e}^-$, only the 0.6$^{\scriptstyle{''}}$ FWHM seeing was simulated. The figure shows the computed central pixel value as a function of source counts, measures of broadening arranged both by source counts and central pixel counts, and properties of the recorded source ellipticity arranged by central pixel counts. The latter two show the systematic changes in PSF parameters (recorded source ellipticity and recorded source rotation angle) that arise from the pixel boundaries that shift in response to the ongoing accumulation of conversions. Not surprisingly, the systematics appear to depend on source integration level, seeing and magnitude of the brighter-fatter effect. 

\begin{figure}[tbph]
\begin{tabular}{rrr}
\includegraphics[trim=1.5cm 1.5cm 1.5cm 1.5cm, clip=true, width=0.25\textwidth, height=0.25\textwidth]{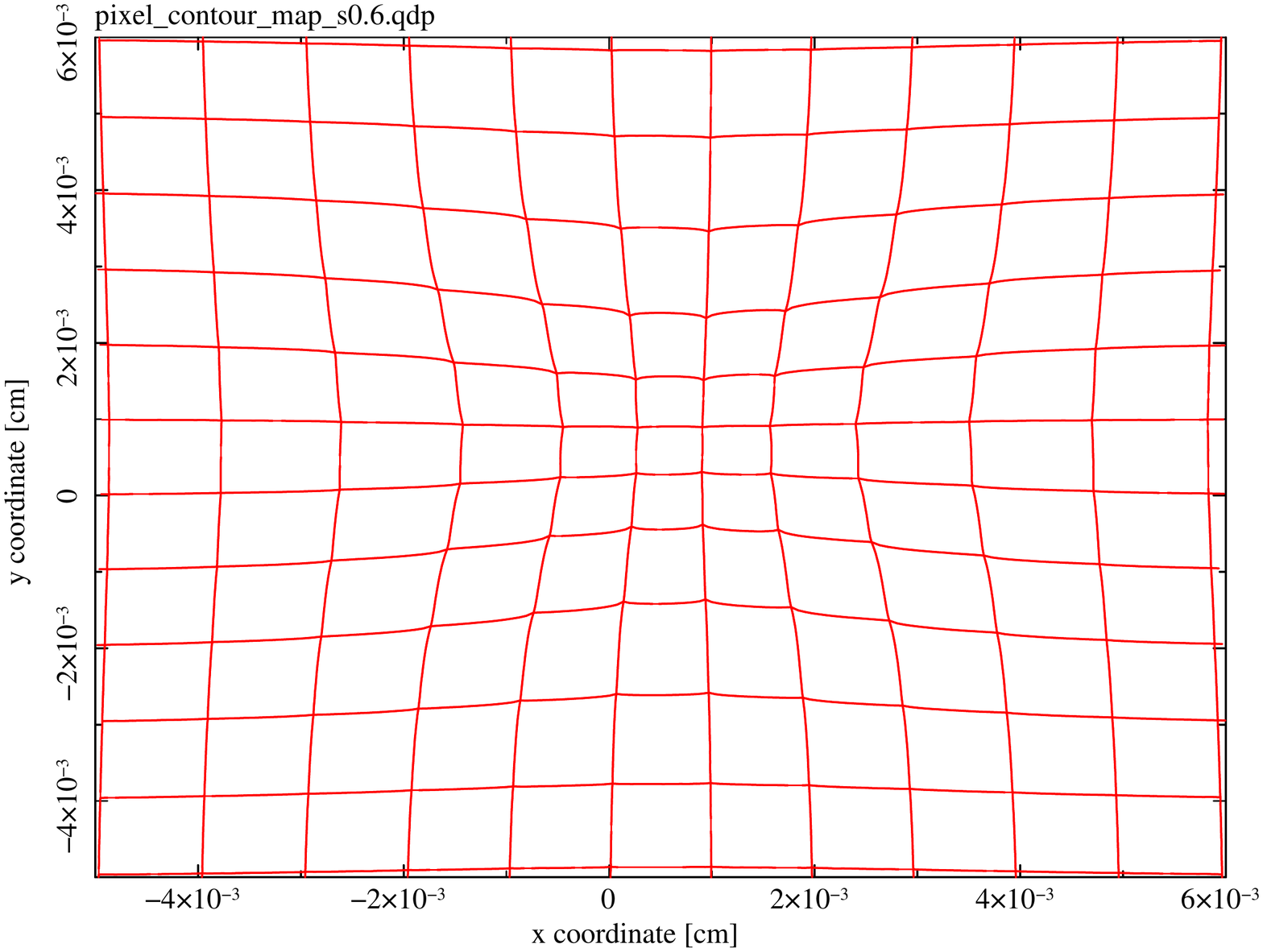}&
\includegraphics[trim=1.5cm 1.5cm 1.5cm 1.5cm, clip=true, width=0.25\textwidth, height=0.25\textwidth]{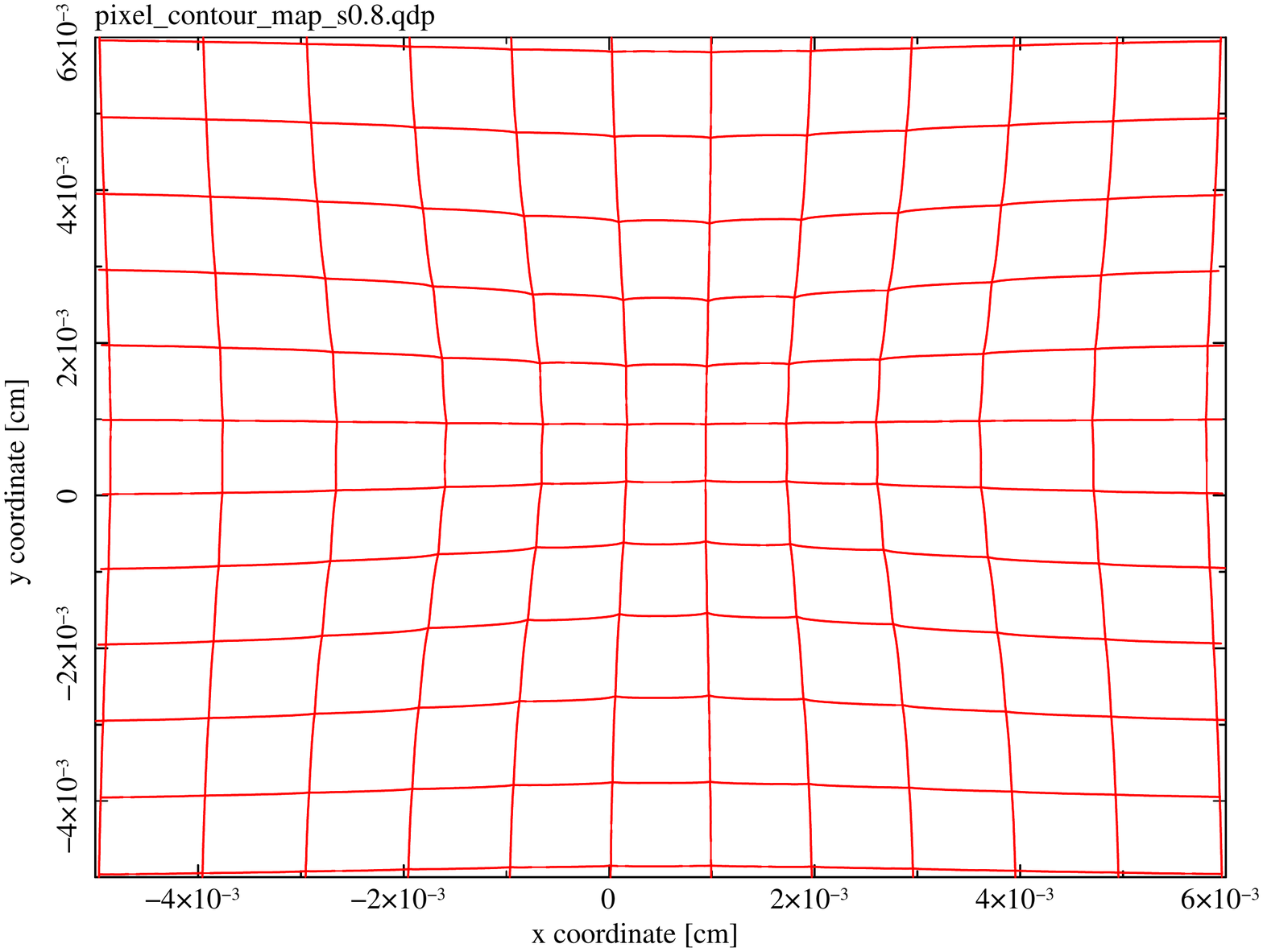}&
\includegraphics[trim=1.5cm 1.5cm 1.5cm 1.5cm, clip=true, width=0.25\textwidth, height=0.25\textwidth]{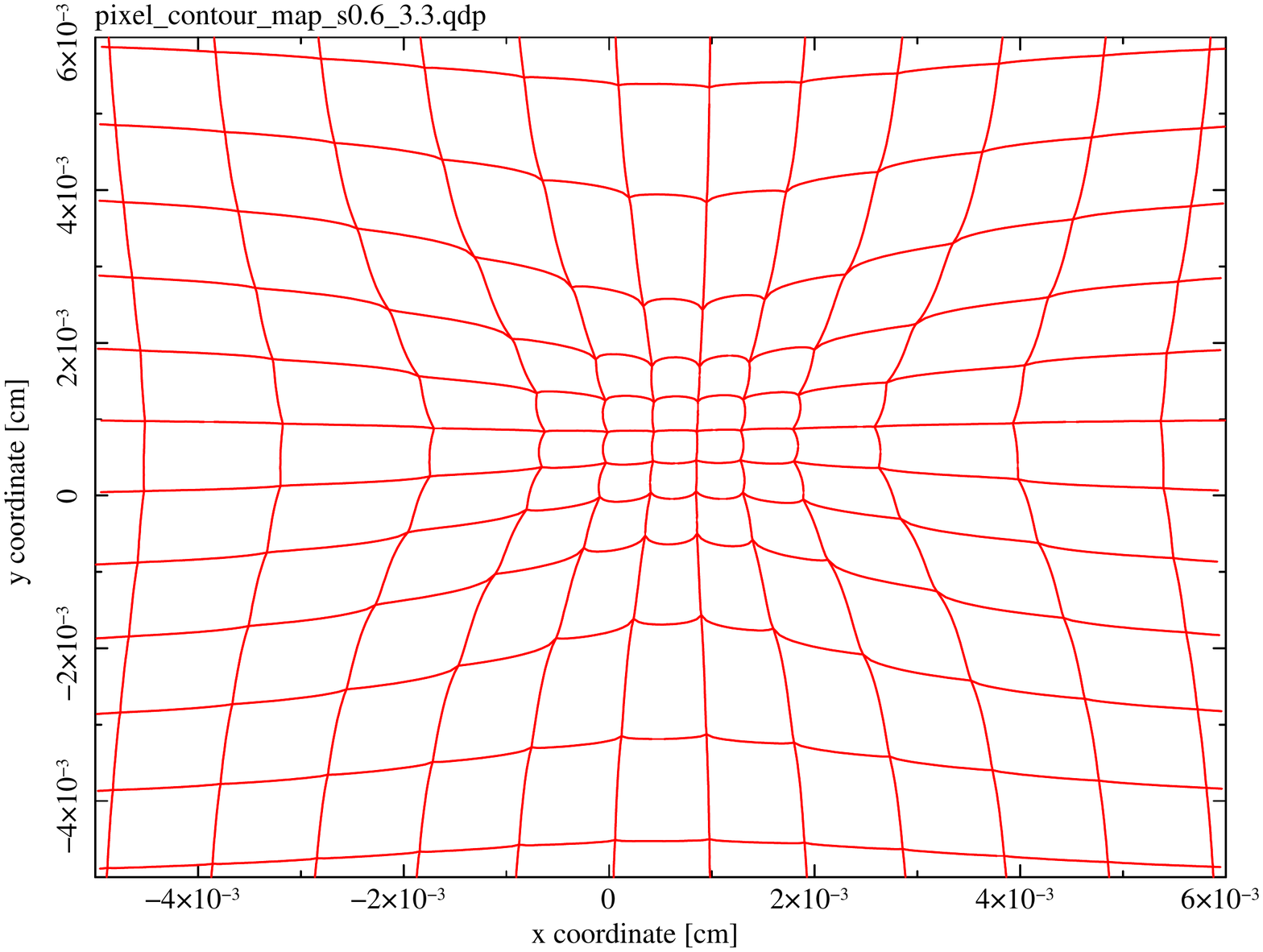}\\
\includegraphics[width=0.28\textwidth, height=0.2\textwidth]{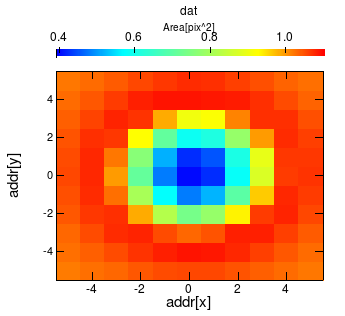}&
\includegraphics[width=0.28\textwidth, height=0.2\textwidth]{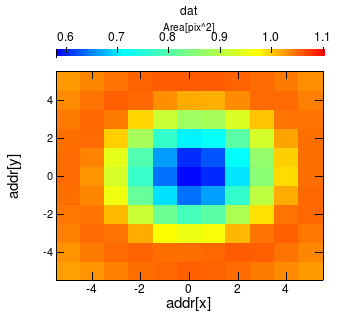}&
\includegraphics[width=0.28\textwidth, height=0.2\textwidth]{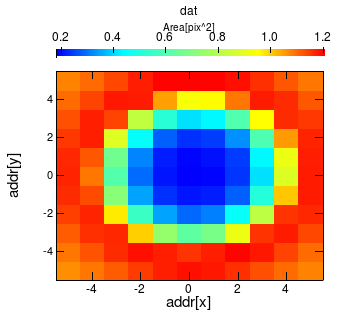}\\
\includegraphics[width=0.28\textwidth, height=0.2\textwidth]{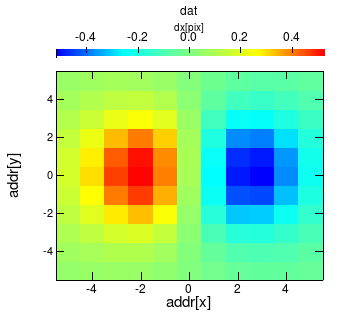}&
\includegraphics[width=0.28\textwidth, height=0.2\textwidth]{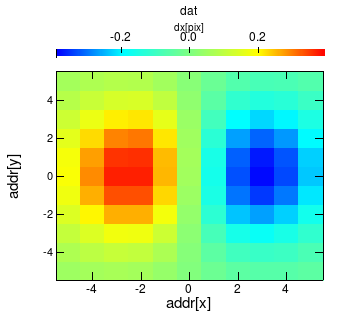}&
\includegraphics[width=0.28\textwidth, height=0.2\textwidth]{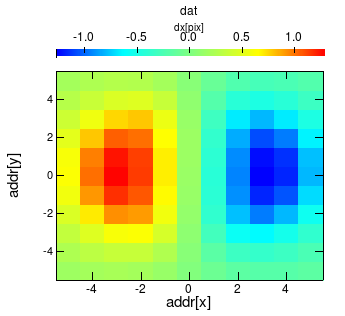}\\
\includegraphics[width=0.28\textwidth, height=0.2\textwidth]{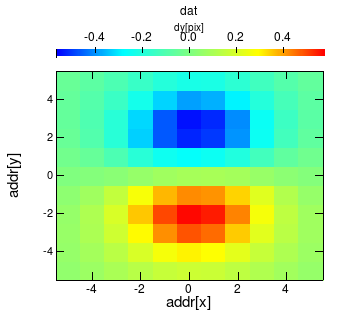}&
\includegraphics[width=0.28\textwidth, height=0.2\textwidth]{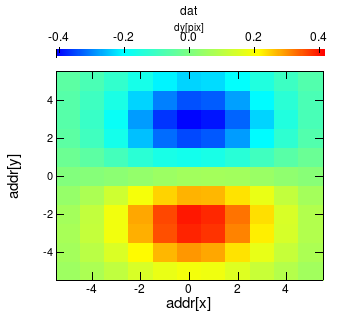}&
\includegraphics[width=0.28\textwidth, height=0.2\textwidth]{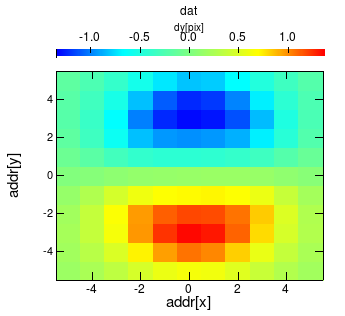}\\
\includegraphics[width=0.28\textwidth, height=0.2\textwidth]{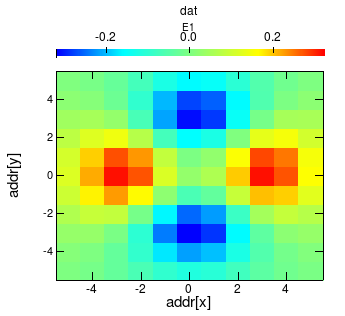}&
\includegraphics[width=0.28\textwidth, height=0.2\textwidth]{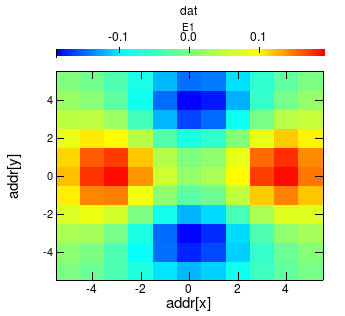}&
\includegraphics[width=0.28\textwidth, height=0.2\textwidth]{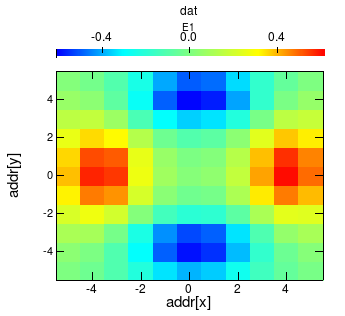}\\
\includegraphics[width=0.28\textwidth, height=0.2\textwidth]{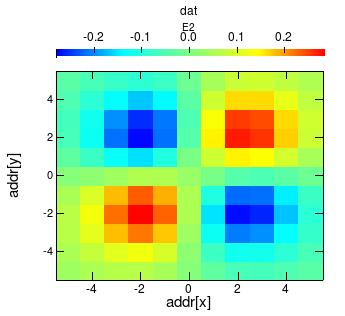}&
\includegraphics[width=0.28\textwidth, height=0.2\textwidth]{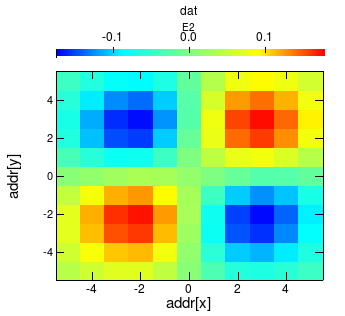}&
\includegraphics[width=0.28\textwidth, height=0.2\textwidth]{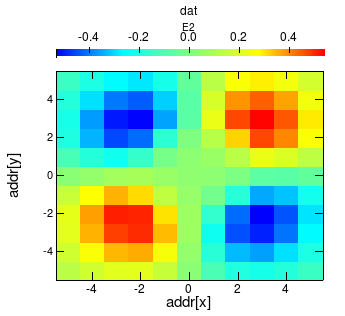}\\
\end{tabular}
\caption{Pixel specific geometric parameters, corresponding to the final, recorded image. Left column: 
0.6$^{\scriptstyle{''}}$ seeing, $\delta\ln A_{00}=-1\times 10^{-5}/\mathrm{e}^-$; center column: 0.8$^{\scriptstyle{''}}$ seeing, $\delta\ln A_{00}=-1\times 10^{-5}/\mathrm{e}^-$; left column: 0.6$^{\scriptstyle{''}}$ seeing, $\delta\ln A_{00}=-3.3\times 10^{-5}/\mathrm{e}^-$. Each column gives the final pixel boundary contours (first row), final pixel area (second row), final pixel astrometric shifts (third and fourth rows), and final pixel ellipticities (fifth and sixth rows). Corrections to individual pixel values may be performed using values from these maps to perform better background subtraction and to estimate source parameters in a way that is less subject to systematics of the brighter-fatter effect.}
\label{fig:bf_recordedpars2}
\end{figure}

If such a brighter-fatter model is validated adequately against available lab data, it is natural to expect that the geometric distortions predicted on a pixel-by-pixel basis can be used in pixel data pipelines to partially cancel the effects of the pixel boundary shifts that apparently occur while the conversions accumulate. The recorded image would be used to determine the pixel boundaries {at the end of the accumulation} using the brighter-fatter template as the Greens function, whereas initial pixel boundaries are assumed to be undistorted. The trajectory connecting the two boundary configurations then depend on details of the flux distribution relative to the undistorted pixel boundary grid and its time history. Figure~\ref{fig:bf_recordedpars2} shows the final configuration for a 15 second illumination by a bright star in $r$ band, AB$\sim$15 magnitudes for the three conditions described above. The maps show the distorted pixel boundary contours, the pixel area distortions $A_{ij}^{\mathrm{final}}$, pixel astrometric distortion $\vec{p}_{ij}^{\mathrm{final}}$ and pixel ellipticity $\vec\epsilon_{ij}^{\mathrm{final}}$, arranged by pixel positions of the recorded image. Notice that accurate background subtraction of the image should also depend to some degree on $A_{ij}^{\mathrm{final}}$. There are a range of choices as to how these maps may be generated and used in pixel data pipelines, and we leave those details for a later discussion.

\begin{figure}[tbp]
\centering
\begin{tabular}{c}
\includegraphics[trim=1.5cm 1.5cm 1.5cm 1.5cm, clip=true, width=0.4\textwidth]{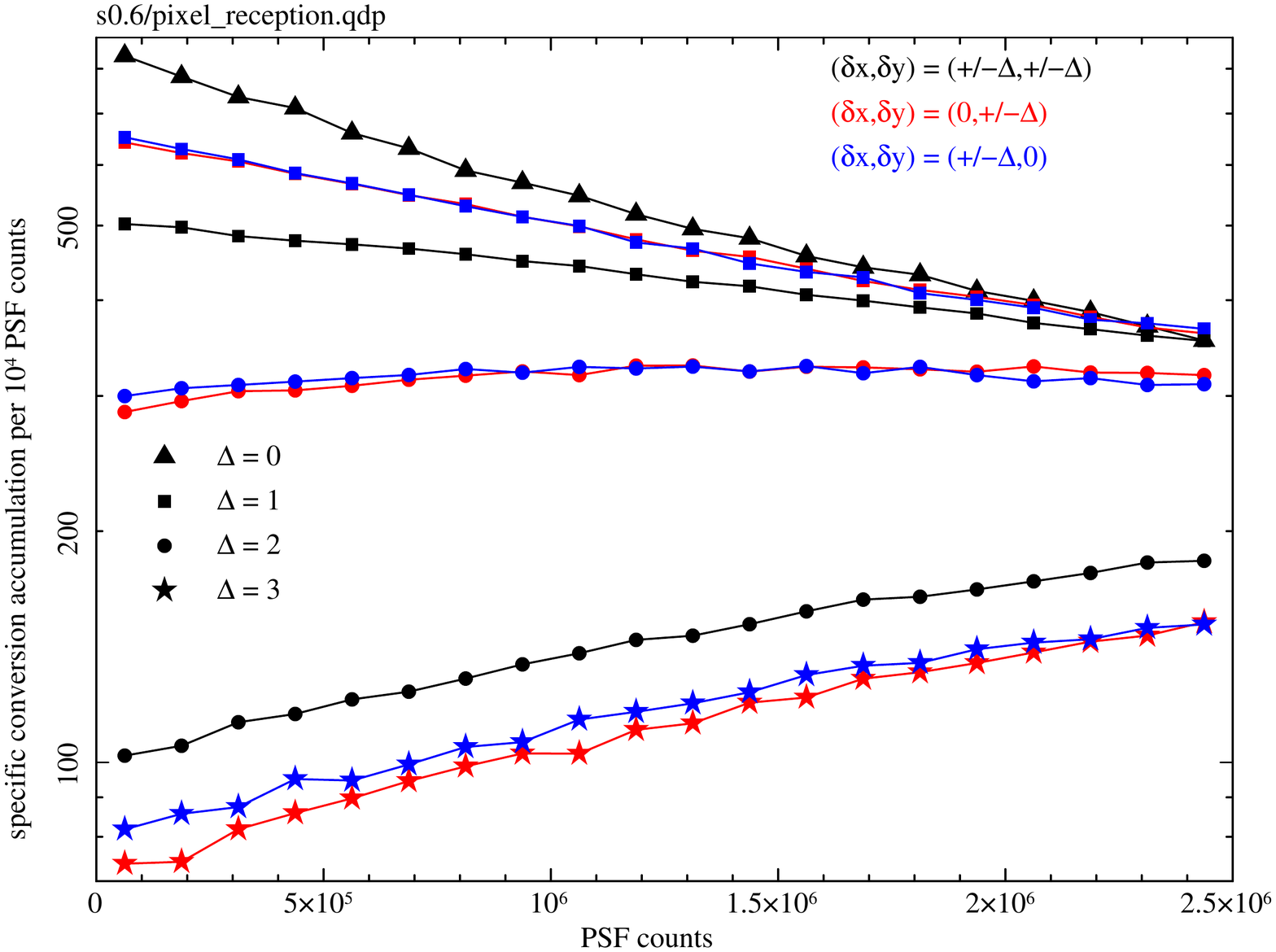}\\
\includegraphics[trim=1.5cm 1.5cm 1.5cm 1.5cm, clip=true, width=0.4\textwidth]{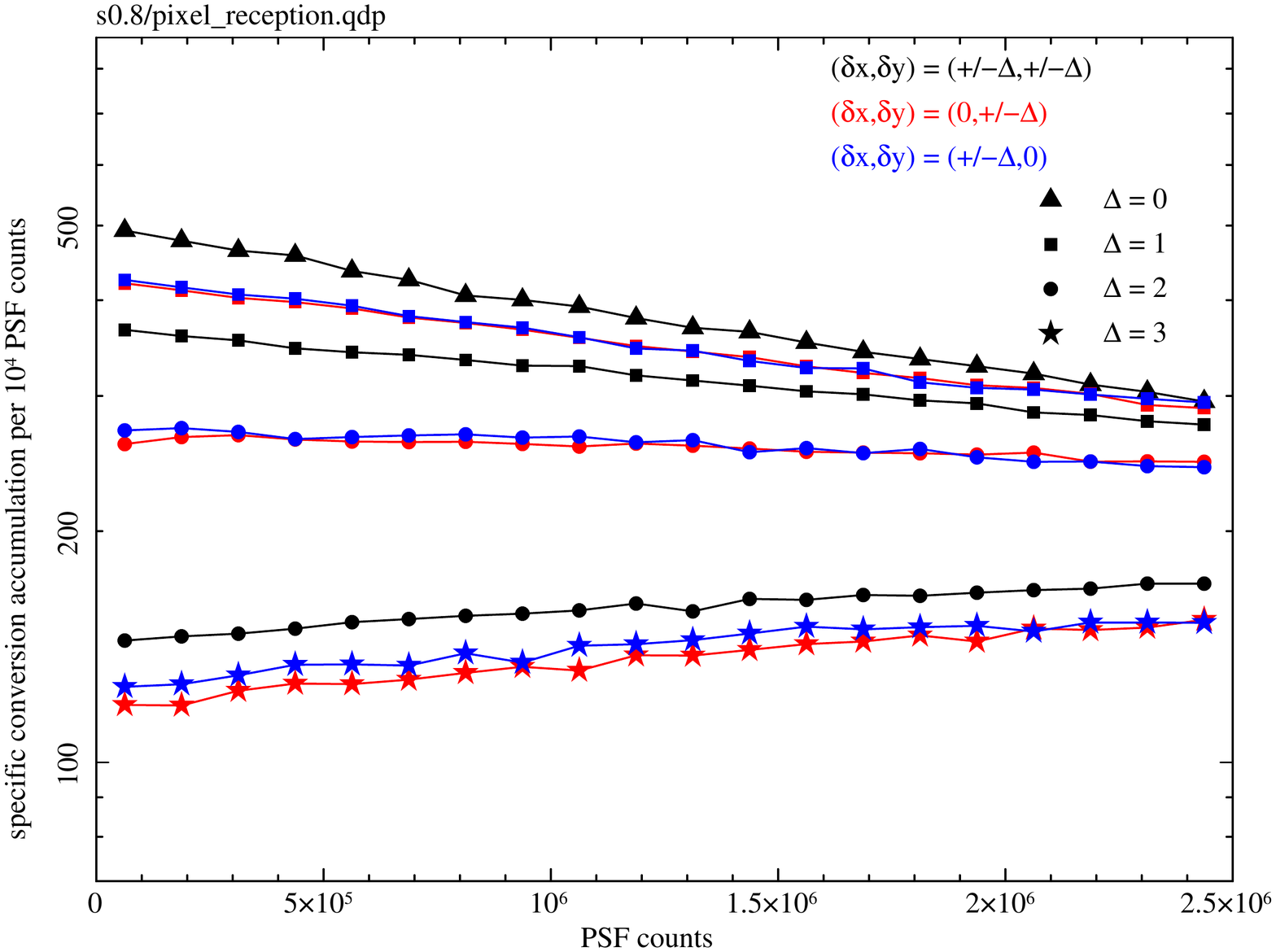}\\
\includegraphics[trim=1.5cm 1.5cm 1.5cm 1.5cm, clip=true, width=0.4\textwidth]{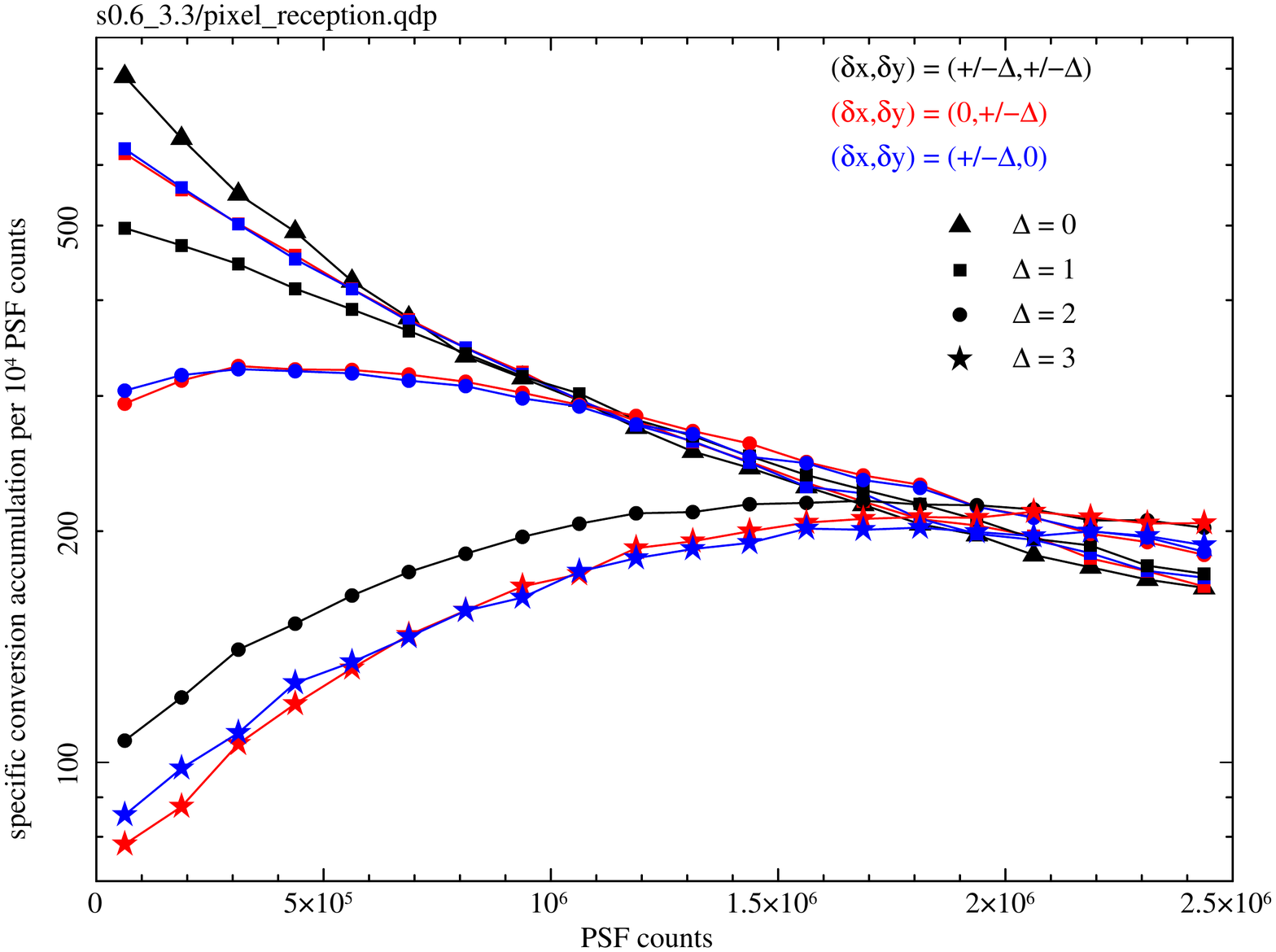}\\
\end{tabular}
\caption{Pixel specific conversion arrival rates (for specific pixels distributed about the PSF centroid), arranged by PSF counts. 
Top:       0.6$^{\scriptstyle{''}}$ seeing, $\delta\ln A_{00}=-1\times 10^{-5}/\mathrm{e}^-$; 
middle:  0.8$^{\scriptstyle{''}}$ seeing, $\delta\ln A_{00}=-1\times 10^{-5}/\mathrm{e}^-$;
bottom:  0.6$^{\scriptstyle{''}}$ seeing, $\delta\ln A_{00}=-3.3\times 10^{-5}/\mathrm{e}^-$.
See text for more information.
}
\label{fig:pixel_reception}
\end{figure}

The unresolved source simulation described here also provides access to instantaneous pixel accumulation rates. Since tracking errors were not included, the only variations in the rates are due to pixel geometric distortions mediated by the brighter-fatter template. Figure~\ref{fig:pixel_reception} shows this for select pixel rate averages in the neighborhood centered on the star. In some extreme cases (e.g., $0.6^{\scriptstyle{''}}$ FWHM, $\delta\ln A_{00}=-3.3\times 10^{-5}/\mathrm{e}^-$, PSF counts~$\sim2\times 10^{6}\,\mathrm{e}^-$) rates at outer pixels overtake the central pixel's rate. In principle, small tracking errors can also lead to image elongations and astrometric biases that scale with source brightness - and these errors should be different along the two pixel address coordinates. Studies that address these effects can be performed with a modest extension of the existing framework.

\section{Conclusions}\label{sec:j}

We have introduced a quantitative modeling framework that can be used to affect a widespread departure from the approximately correct assumptions that pixel boundaries remain, over the dynamic ranges of interest, fixed, independent of sensor position and local illumination history. Rich datasets can be used to tune parameters of this drift model and be tested against validation requirements. Results of the tuned model are portable and can be applied in superposition outside of the sensor drift and partition calculator. We consider how the overall partition model may be broken up into influence components that act on longer length scales (and should have moderate chromatic and backside bias voltage dependences), and influence components that act on shorter length scales, that have significantly reduced dependences on those variables. The superposition of brighter-fatter templates on top of effects that inject bias ({\it e.g.}, edge rollof, tracking errors) - should produce brightness-correlated astrometric and shape transfer errors that should be examined.

\acknowledgments
We enjoyed insightful discussions with Pierre Antilogus, Robert Lupton, Chris Stubbs, and Daniel Gruen. LSST project activities are supported in part by a Cooperative Agreement with the N.S.F. managed by A.U.R.A., and the D.O.E. Additional LSST funding comes from private donations, grants to universities, and in kind support from LSSTC Institutional Members.
\bibliographystyle{JHEP}
\bibliography{precision_astronomy_arasmus_2014_jinst}

\end{document}